\documentclass[aps,prd,reprint,preprintnumbers,nofootinbib,superscriptaddress]{revtex4-1}
\pdfoutput=1
\usepackage{natbib}
\usepackage{amsmath}
\usepackage{amssymb}
\usepackage{amsfonts}
\usepackage{hyperref}
\usepackage{epsfig}
\usepackage{graphicx}
\usepackage{slashed} 
\usepackage{mathrsfs} 
\usepackage{subcaption}
\allowdisplaybreaks
\usepackage{url}
\usepackage{color}
\definecolor{myred}{rgb}{0.6,0,0} 
\definecolor{myblue}{rgb}{0,0.2,0.4}
\definecolor{mygreen}{rgb}{0,0.9,0.1}
\definecolor{hc}{rgb}{.9,0.1,0.7}
\definecolor{hcout}{rgb}{.9,0.7,0.9}
\definecolor{Orange}{rgb}{1.,0.65,0.}
\makeatletter
\newcommand{\fmslash}[2][0mu]{%
  \mathchoice
    {\fmsl@sh\displaystyle{#1}{#2}}%
    {\fmsl@sh\textstyle{#1}{#2}}%
    {\fmsl@sh\scriptstyle{#1}{#2}}%
    {\fmsl@sh\scriptscriptstyle{#1}{#2}}}
\newcommand{\fmsl@sh}[3]{%
  \m@th\ooalign{$\hfil#1\mkern#2/\hfil$\crcr$#1#3$}}
\makeatother
\newcommand{\lsim}{{\;\raise0.3ex\hbox{$<$\kern-0.75em\raise-1.1ex\hbox{$\sim$}}\;}}
\newcommand{\gsim}{{\;\raise0.3ex\hbox{$>$\kern-0.75em\raise-1.1ex\hbox{$\sim$}}\;}}
 
\usepackage{array}
\newcolumntype{C}[1]{>{\centering\arraybackslash$}p{#1}<{$}}
\usepackage{bm} 
\newcommand{\be}{\begin{equation}}
\newcommand{\ee}{\end{equation}}
\newcommand{\bes}{\begin{equation*}}
\newcommand{\ees}{\end{equation*}}
\newcommand{\bea}{\begin{eqnarray}}
\newcommand{\eea}{\end{eqnarray}}
\newcommand{\beas}{\begin{eqnarray*}}
\newcommand{\eeas}{\end{eqnarray*}}

\begin{document}

\preprint {HRI-RECAPP-2019-004}

\title{Heavy dark matter particle annihilation in dwarf spheroidal galaxies: radio
signals at the SKA telescope}

\author{Arpan Kar}
\email{arpankar@hri.res.in}
\affiliation{Regional Centre for Accelerator-based Particle Physics, Harish-Chandra Research Institute, HBNI, Chhatnag Road, Jhunsi, Allahabad - 211 019, India}
\author{Sourav Mitra}
\affiliation{Surendranath College, 24/2 M. G. ROAD, Kolkata, West Bengal 700009, India}

\author{Biswarup Mukhopadhyaya}
\altaffiliation[Present affiliation: ]{Indian Institute of Science Education and Research Kolkata, Mohanpur, West Bengal 741246, India}
\affiliation{Regional Centre for Accelerator-based Particle Physics, Harish-Chandra Research Institute, HBNI, Chhatnag Road, Jhunsi, Allahabad - 211 019, India}

\author{Tirthankar Roy Choudhury}
\affiliation{National Centre for Radio Astrophysics, TIFR, Post Bag 3, Ganeshkhind, Pune 411007, India}

\begin{abstract}
A weakly interacting dark matter candidate is difficult to
detect at high-energy colliders like the LHC, if its mass is
close to, or higher than a TeV. On the other hand, pair-annihilation of
such particles may give rise to $e^+ e^-$ pairs in dwarf spheroidal
galaxies (dSph), which in turn can lead to radio synchrotron signals
that are detectable at the upcoming Square Kilometre Array (SKA) telescope
within a moderate observation time. We investigate the circumstances
under which this complementarity between collider and radio signals
of dark matter can be useful in probing physics beyond the standard
model of elementary particles. Both particle physics issues and the roles
of diffusion and electromagnetic energy loss of the $e^\pm$ are taken into
account. First, the criteria for detectability of trans-TeV dark matter are
analysed independently of the particle physics model(s) involved.
We thereafter use some benchmarks based on a popular scenario, namely,
the minimal supersymmetric standard model. It is thus shown that the radio
flux from a dSph like Draco should be observable in about 100 hours at
the SKA, for dark matter masses upto 4-8 TeV. In addition, the regions in the space spanned
by astrophysical parameters, for which such signals should be detectable at the SKA,
are marked out.
\end{abstract}

\maketitle

\section{Introduction}\label{Introduction}

Some yet unseen weakly interacting massive particles (WIMP) are often
thought of as constituents of  dark matter (DM) in our universe. Many scenarios 
beyond the standard model (SM) of particle physics including them are regularly 
proposed, with various phenomenological implications.
They are constrained by the data on relic density \cite{Ade:2013zuv} of the universe as well as
various direct search experiments \cite{Aprile:2017iyp,PhysRevLett.118.251302}. It is  expected that a WIMP DM candidate
should also be detected at the Large Hadron Collider (LHC) in the form of 
missing transverse energy (MET) (see for example \cite{Carena:2019pwq,Ellis:2018zpk,Abe:2018bpo}).

Detectability at colliders and also in direct search experiments,
however, depends on the mass as well as the interaction cross-section
of the DM particle. In particular, detection becomes rather difficult if the
WIMP mass approaches a TeV \cite{SHCHUTSKA2016656}. 
For a weakly interacting DM, production at the LHC has to depend mostly on Drell-Yan
processes where the rate gets suppressed by the $s$, the square of the subprocess centre-of-mass
energy, and also by the parton distribution function at high $x$. An exception is the 
minimal supersymmetric standard model (MSSM) \cite{Martin:1997ns}; there the production of colored superparticles (squarks/gluions) via strong interaction may be followed by decays in cascade leading to DM pair-production. However, here, too, the event rates go down drastically when one looks at the current constraints on colored superparticle masses as well as the various
other limits on the MSSM spectrum.
On the whole, while a near-TeV DM particle still admits of  
some hope at the LHC in the MSSM \cite{SHCHUTSKA2016656,ATL-PHYS-PUB-2014-010,CMS-PAS-FTR-13-014,ATL-PHYS-PUB-2013-011}, the reach is considerably lower for most other scenarios where the
`dark sector' is at most weakly interacting \cite{Suematsu:2007dc,Morgante:2017kuo}. It is therefore a challenge to think of
 additional indirect evidence, if a trans-TeV DM particle has to be explored.
 
 The annihilation of DM particles in our galaxy as well as in extra-galactic objects
 leads to gamma-ray signals \cite{Daylan:2014rsa,Calore:2014nla,Kar:2017oer,Kar:2017dmg,Karwin:2016tsw,Beck:2017wxu,Achterbeg:2015dca} 
 as well as positrons, antiprotons etc \cite{HEKTOR201458,PhysRevD.97.115012,Cholis:2019ejx,Cui:2018klo}. Constraints
 have been imposed on DM annihilation rates in 
 various ways out of the (non)-observation of such signals \cite{TheFermi-LAT:2017vmf,Fermi-LAT:2016uux,PhysRevLett.115.231301,Cuoco:2017iax}. 
 An alternative 
 avenue to explore is that opened by radio synchrotron emission from
 galaxies, arising out of electron-positron pairs generated from
 DM annihilation \cite{Regis:2017oet,Natarajan:2015hma,Natarajan:2013dsa,Spekkens:2013ik}.  
 In this paper, we focus on the potential of the upcoming 
 Square Kilometre  Array (SKA) radio telescope \cite{SKA} in this regard.   
 
 While the prospect of radio fluxes unveiling DM annihilation has been explored
 in earlier works \cite{Regis:2017oet,Natarajan:2015hma,Natarajan:2013dsa,Spekkens:2013ik,Colafrancesco:2015ola,Cirelli:2016mrc, Regis:2014tga}, it was pointed out in 
 reference \cite{Kar:2019mcq} that SKA opens up a rather
 striking possibility. The annihilation of trans-TeV DM pairs in dwarf spheroidal 
 galaxies (dSph) lead to electron-positron pairs which, upon acceleration by the 
 galactic magnetic field, produces such radio synchrotron emission.  dSph's are
 suitable for studying DM, since star formation rates there are low \cite{Strigari:2008ib,Geha:2008zr}, thus
 minimising the possibility of signal originating from astrophysical processes.
 Their generic faintness prompts one to concentrate on such galaxies which are
 satellites of the Milky Way. The SKA can ensure sufficient 
 sensitivity required to detect the faint signal from the sources,
 and at the same time, will have high enough resolution to
 remove the foregrounds. It was
 shown in \cite{Kar:2019mcq} that about 100 hours of observation at the SKA can take us
 above the detectability  threshold for radio signals from the 
 annihilation of DM particles in the 5-10 $TeV$ range. Of course, the compatibility
 of such massive WIMP with the observed relic density requires a dark 
 sector spectrum with enough scope for co-annihilation in early universe,
 as was demonstrated  in \cite{Kar:2019mcq} in the context of the MSSM. What one learns
 from such an exercise is that the probe of at least some DM scenarios should 
 thus be possible on a time scale comparable with the running period of
 the LHC, and that the reach of the LHC \cite{SHCHUTSKA2016656,ATL-PHYS-PUB-2014-010,CMS-PAS-FTR-13-014,ATL-PHYS-PUB-2013-011} 
 for WIMP detection may be
 exceeded considerably through such a probe. It was also found that cases
 where SKA could observe radio fluxes from a dSph were consistent with
 limits from $\gamma$-ray observations as well as antiparticles in cosmic rays.

In order to ascertain which scenarios are more accessible in such radio probes,
one needs to understand in detail the mechanisms whereby high-mass DM 
particles can produce higher radio fluxes, and also the effects of 
astrophysical processes that inevitably affect radio emission. This is the 
task undertaken in the present work. 
 
The spectrum as well as the dynamics of the particle physics scenario,
along with the DM profile in a dSph, is responsible for DM annihilation 
as well as the subsequent cascades leading to electron-positron pairs.\footnote{In principle, electron-positron pairs may also be directly produced in a complete model-independent scenario. We have referred to cascade since our benchmark correspond to $b\bar{b}, t\bar{t}, 
W^+W^-$ and $\tau^+\tau^-$ as dominant annihilation channels, inspired by the MSSM where direct 
$e^+e^-$ is not found to dominate for high DM masses.}
The electron(positron) energy distribution at the source function level is also the 
determined by the above factors \cite{Colafrancesco:2005ji,McDaniel:2017ppt,Linden:2010eu}.  However, they subsequently pass
through the interstellar medium (ISM) of the galaxy, facing several additional effects.
These include diffusion as well as electromagnetic energy loss in various
forms, including Inverse Compton effect, Synchrotron loss, Coulomb effect
and Bremsstrahlung effect \cite{Colafrancesco:2005ji,Beck:2015rna,Colafrancesco:2006he,McDaniel:2017ppt}. 
Besides, the galactic magnetic field is operative all along.  The way these affect 
the final $e^+ e^-$ energy distribution is  highly inter-connected and nonlinear.
For example, while the magnetic field causes electrons to lose more energy
in synchrotron radiation, it puts a check on the reduction of the flux 
through diffusion by 
confining them longer within the periphery of the dSph. 
This check applies to electrons at lower energies
at the cost of those at higher energies. Also, electromagnetic
energy loss  enhances the population of low-energy electrons and positrons,
the enhancement being more when they have higher kinematic limits,
enabled by higher mass of the DM particle. Such low-energy $e^+ e^-$
pairs enhance the flux in the frequency range appropriate for a radio
telescope. 

In the following sections, we analyse the various ingredients in the radio 
flux generation process, as outlined above. We first do this for fixed DM particle
masses, and for values of their annihilation cross-sections fixed by hand. The
relative strengths of the effects of  the particle theoretical scenario as
well as diffusion and  radiative process  are thus assessed. This also
serves to evolve an understanding of the dependence on the diffusion
coefficient, parameters involved in electromagnetic energy loss, and, of
course, the strength of the galactic magnetic field. If the nature of the DM
particle(s) is known in independent channels,   then the observation of the
observed radio flux may be turned around to improve our understanding of these
astrophysical parameters, using the results presented here.

Finally, we use some theoretical benchmark points to demonstrate
the usefulness of the SKA in probing trans-TeV DM. A few sample MSSM 
spectra are  used, largely because they offer the scope of 
co-annihilation that is so essential for maintaining the right relic density \cite{Kar:2019mcq}.
Using the minimum annihilation cross-section  required
for any DM mass for detection at SKA (in 100 hours), and the
maximum value of this cross section that is compatible with limits from
$\gamma$-ray and cosmic-ray data, we show that several 
benchmark points with DM mass in the 1-8 $TeV$ range, 
which are yet to be ruled out by any observation can be investigated 
in 100 hours of SKA observation.

We have organised the paper as follows. In section \ref{Essential Processes} we 
have discussed in somewhat brief manner the calculations of synchrotron flux
originating from DM annihilation inside a dSph. In section \ref{astrophysics effect}
we have analysed the effects of various astrophysical parameters in different
steps of the production of radio flux. Section \ref{Effect of Heavy DM} describes
the features of heavy DM. In section \ref{Detectability Curves} we have
shown the detectability curves or threshold limits for observing radio flux in SKA,
in both model independent as well as model dependent way. 
We have also shown the final radio fluxes for some theoretical benchmarks.
Finally in section \ref{Conclusion} we conclude.


\section{Essential Processes: Recapitulation}\label{Essential Processes}

We start with a brief resume of sequence of processes that leads to radio
flux from a dSph, as discussed, for example, in 
\cite{Natarajan:2015hma,Natarajan:2013dsa,Colafrancesco:2005ji,Beck:2015rna,Colafrancesco:2006he,McDaniel:2017ppt}.
Dark matter pair annihilation inside of a dSph can produce SM particle pairs such as 
$b\bar{b}, \tau^+\tau^-, W^+W^-, t\bar{t}$ etc. These particles then cascade and give rise to
large amount of $e^\pm$ flux. The resulting energy distribution of which
can be obtained from the source function \cite{McDaniel:2017ppt,Linden:2010eu}:

\begin{equation}
Q_e(E,r) = \langle \sigma v \rangle \left\lbrace \sum_{f} {\frac{dN^e_f(E)}{dE}} B_f\right\rbrace   N_{pairs} (r)
\label{source_function}
\end{equation}
where $\langle \sigma v \rangle$ and $N_{pairs} (r)
\left(= \frac{\rho^2_{\chi}(r)}{2 m_{\chi}^2}  \right)$ are respectively 
the velocity averaged
annihilation rate and the number 
density of DM pairs  inside the dSph. $m_{\chi}$ is the DM
mass and $\rho_{\chi}(r)$ is the DM density profile in the dSph
as a function of radial distance $r$ from the centre of the 
dSph. For our analysis 
we have taken Draco\footnote{In this paper we have used the dSph Draco to illustrate
our points, primarily because the relevant parameters such as the $J$-factor are
somewhat better constrained for this object \cite{Geringer-Sameth:2014yza}. 
However, similar conclusions apply to other
dSph's such as Seg1, Carina, Fornax, Sculptor etc. \cite{Geringer-Sameth:2014yza,Choquette:2017nqk,Kar:2019hnj}}
dSph assuming an Navarro-Frenk-White (NFW) profile \cite{Navarro:1995iw}, 
with

\begin{equation} 
\rho_{\chi} (r) = \frac{\rho_s}{\left( \frac{r}{r_s} \right)  
\left( 1 + \frac{r}{r_s} \right)^2 }
\label{rho_NFW}
\end{equation}
where $\rho_s = 1.4$ $GeV. cm^{-3}$ and $r_s = 1.0$ $kpc$ \cite{Colafrancesco:2006he,McDaniel:2017ppt}. $\frac{dN^e_f(E)}{dE} B_f$ estimates 
the number of $e^\pm$ produced with energy $E$ per annihilation
in any of the aforementioned SM channel 
($f$) having branching fraction $B_f$.

\begin{figure*}
\includegraphics[height=0.32\textwidth, angle=0]{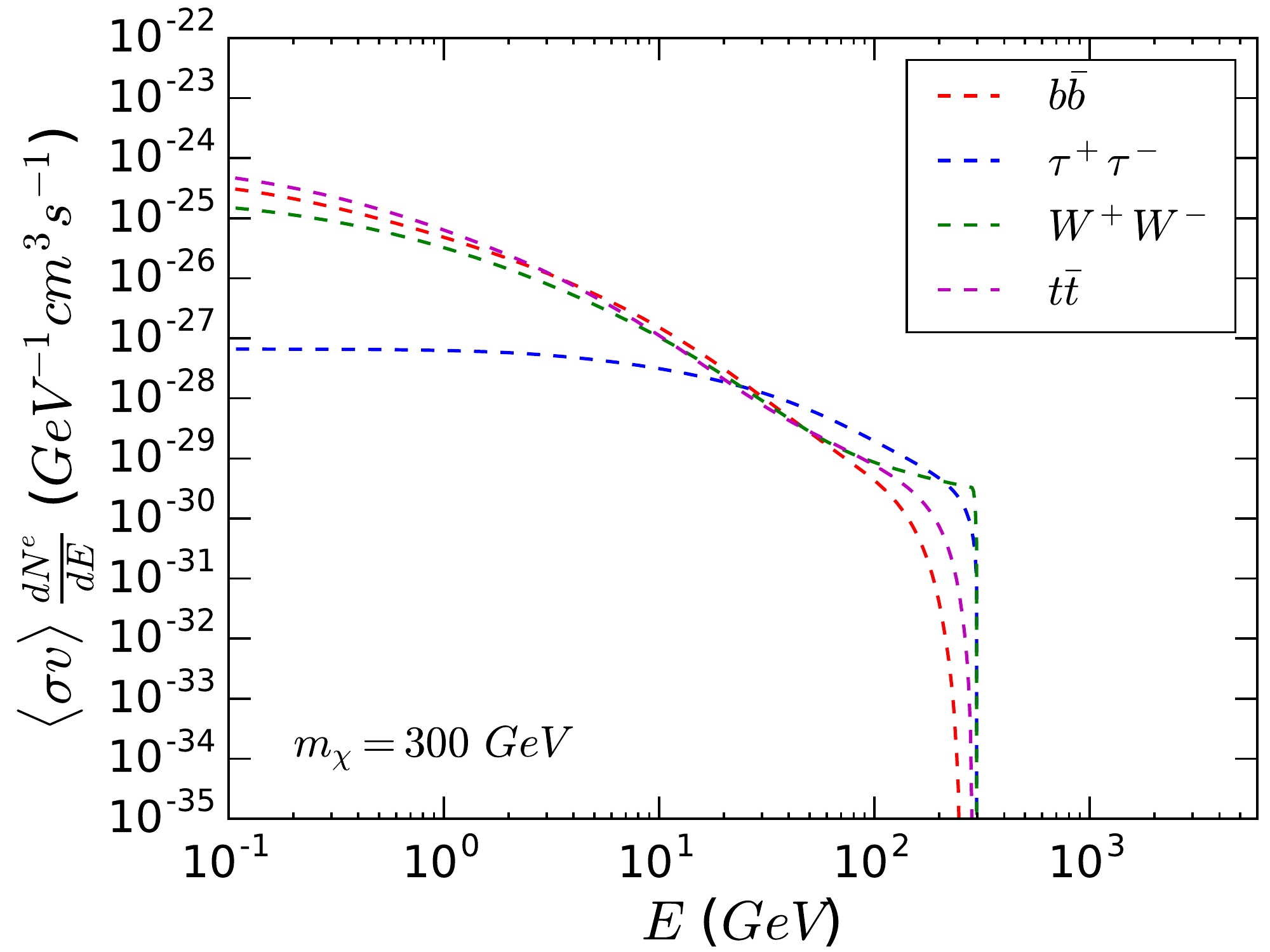}\hspace{3mm}%
\includegraphics[height=0.32\textwidth, angle=0]{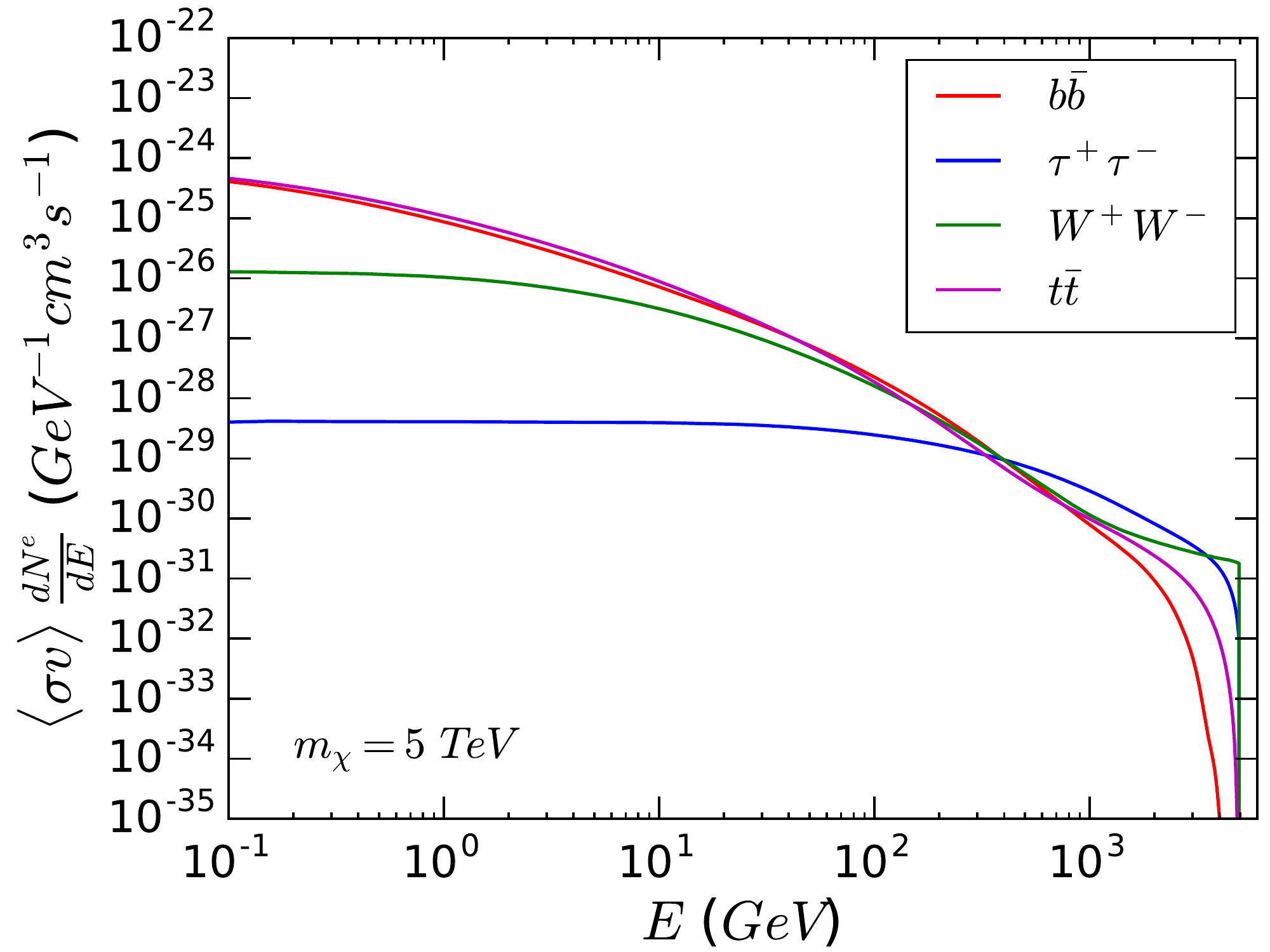}
\includegraphics[height=0.32\textwidth, angle=0]{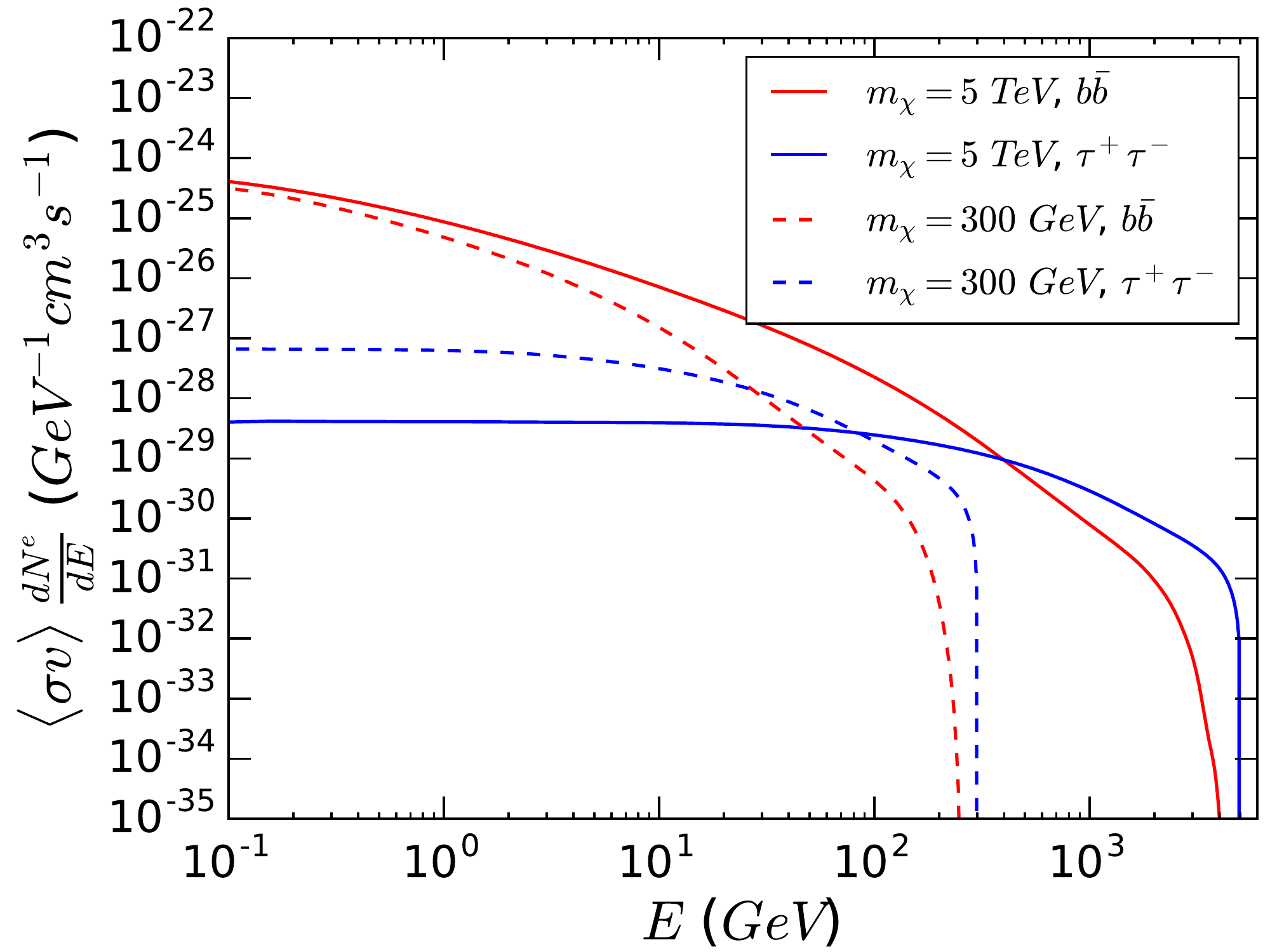}
\caption{{\it Upper panel:} Source functions per unit
annihilation $(\langle \sigma v \rangle$ $\frac{dN^e}{dE})$ vs. 
electron energy ($E$) in different annihilation channels for two DM masses, 300 $GeV$ 
(upper left panel) and 5 $TeV$ (upper right panel). 
{\it Lower panel:} Comparison of the same for two DM masses, 300 $GeV$ (dashed curves) and 
5 $TeV$ (solid curves) in two annihilation channels, $b\bar{b}$ (red lines) and 
$\tau^+\tau^-$ (blue lines).
Annihilation rate for each panel is 
$\langle \sigma v\rangle$ = $10^{-26} \mbox{cm}^3 \mbox{s}^{-1}$.}
\label{Q_E}  
\end{figure*}

Figure \ref{Q_E} shows the electron energy distribution for four different annihilation channels ($b\bar{b}, \tau^+\tau^-, W^+W^-, t\bar{t}$) and 
for two DM masses, namely, $m_{\chi} =$ 300 $GeV$ ({\it upper left panel}) 
5 $TeV$ ({\it upper right}).  
For all cases $\langle \sigma v \rangle$ has been assigned a fixed value ($10^{-26} \mbox{cm}^3 \mbox{s}^{-1}$), 
and the branching fraction corresponding to each channel has been set in turn
at $100\%$. In addition, a comparison between cases with the two masses has also been
shown in the lower panel  for the $b\bar{b}$ and $\tau^+ \tau^-$ channel. 
The prediction for  $t\bar{t}$ and $W^+ W^-$ falls in between those curves. 
All these energy distributions are obtained using micrOMEGAs \cite{micrOMEGAs,Belanger:2010gh}. 
Further discussions on these curves will be taken up in section
\ref{Effect of Heavy DM}.

The electrons produced in DM pair annihilation, diffuse through the 
interstellar medium (ISM) in the galaxy and lose energy through various electromagnetic processes. 
Assuming a steady state and 
homogeneous diffusion, the equilibrium distribution $\frac{dn}{dE}$ can be 
obtained by solving the
equation \cite{Natarajan:2013dsa,Colafrancesco:2005ji,Beck:2015rna,Colafrancesco:2006he,McDaniel:2017ppt,Storm:2016bfw}

\begin{equation} 
D(E) \nabla^2 \left(\frac{dn}{dE}\right) +
\frac{\partial}{\partial E}\left(b(E) \frac{dn}{dE}\right) +
Q_e(E,r) = 0
\label{transport1}
\end{equation}
where $D(E)$ is the diffusion parameter. The exact form of $D(E)$ is not known
for dSphs, hence for simplicity, we assume it to have a Kolmogorov spectrum $D(E) = D_0 \left(\frac{E}{GeV}\right)^{0.3}$, where $D_0$ is the diffusion coefficient \cite{Colafrancesco:2006he,McDaniel:2017ppt,Spekkens:2013ik}. 
Very little is known about the value of $D_0$ for dSphs
due to their very low luminosity.
A choice like $D_0 = 3 \times 10^{28} \mbox{cm}^2 \mbox{s}^{-1}$
is by and large reasonable for a dSph such as Draco 
\cite{Colafrancesco:2006he}, although higher values of $D_0$, too, are sometimes
used as benchmark \cite{Regis:2014tga}. 
We have mostly used $D_0 = 3 \times 10^{28} \mbox{cm}^2 \mbox{s}^{-1}$, and demonstrated side by side the effect of other values of this largely unknown 
parameter.
As explained in \cite{Natarajan:2015hma,Natarajan:2013dsa,Jeltema:2008ax}, 
assuming a proper scaling, $D_0$ for a dSph can also in principle have
a value one order lower than that in the Milky Way \cite{Donato:2003xg}.
However, the present study is not restricted to this value, mainly 
$D_0 \lesssim 3 \times 10^{28} \mbox{cm}^2 \mbox{s}^{-1}$. The detectability of the
radio flux for higher $D_0$, too, has been studied by us, as will be seen when we
come to Figure \ref{B_D0}.

The energy loss term $b(E)$
takes into account all the electromagnetic
energy loss processes such as Inverse Compton (IC) effect, 
Synchrotron (Synch) radiation, Coulomb loss (Coul), bremsstrahlung (Brem) etc. 
and their combined effect can be expressed as \cite{Colafrancesco:2005ji,Beck:2015rna,McDaniel:2017ppt}
 
\begin{eqnarray} 
b(E) &=& b^0_{IC} \left(\frac{E}{GeV}\right)^2 + b^0_{Synch} \left(\frac{E}{GeV}\right)^2 \left(\frac{B}{\mu G}\right)^2 \nonumber\\
&& + b^0_{Coul} n_e \left[1 + \frac{\log\left(\frac{E/m_e}{n_e}\right)}{75} \right]\nonumber \\
&& + b^0_{Brem} n_e \left[\log\left(\frac{E/m_e}{n_e}\right) + 0.36 \right]
\label{energy_loss}
\end{eqnarray}
where $m_e$ is the electron mass and $n_e$ is the average thermal electron density in the dSph 
(value of $n_e \approx 10^{-6}$). 
Values of the energy loss coefficients are taken to be 
$b^0_{IC} \simeq 0.25$, $b^0_{Synch} \simeq 0.0254$, $b^0_{Coul} \simeq 6.13$
$b^0_{Brem} \simeq 1.51$, all in units of $10^{-16}$ $GeV s^{-1}$
\cite{Colafrancesco:2005ji,Beck:2015rna,McDaniel:2017ppt}.
In a dSph like Draco, the first two terms (i.e. IC and Synch) in 
Equation \ref{energy_loss} dominates
over the last two terms (i.e. Coul and Brem) for $E > 1$ $GeV$ \cite{Natarajan:2013dsa}.
The energy loss term $b(E)$ depends on the galactic magnetic field ($B$) through the 
synchrotron loss term which goes as $B^2$.
It is extremely difficult to get insights (say, through polarization experiments) for the magnetic field properties of dSph. The lack of
any strong observational evidence suggests that the
magnetic fields could be, in principle, extremely low. On
the other hand, there could be various effects which can
significantly contribute to the magnetic field strengths in dSph.
In fact, numerous theoretical arguments have been proposed for values of
$B$ at the $\mu G$ level. 
It has been argued, for example, that the observed fall of $B$ from the centre
of Milky Way to its peripheral region can be linearly extrapolated to nearby dSph's,
leading to $B \gtrsim 1$ $\mu G$. The possibility of the own magnetic field of a dSph
has also been suggested. 
For detailed discussions we refer the reader
to reference \cite{Regis:2014koa}.
For the dSph considered in this work, we have mostly used $B = 1$ $\mu G$
\cite{Colafrancesco:2006he,McDaniel:2017ppt,Spekkens:2013ik}, but
more conservative values like $B \sim 10^{-1} - 10^{-3}$ $\mu G$ have also been considered.  
The manner in which diffusion and electromagnetic processes affect our observables
will be discussed in detail in sections \ref{astrophysics effect} and \ref{Detectability Curves}.
The overall potential of the SKA in probing regions in the $D_0 - B$ space in a correlated
fashion will be reported in section \ref{Detectability Curves}.

Equation \ref{transport1} has a solution of the form \cite{Natarajan:2013dsa,Colafrancesco:2005ji,Colafrancesco:2006he,McDaniel:2017ppt}

\begin{equation} 
\frac{dn}{dE} (r,E) = \frac{1}{b(E)} \int_E^{m_{\chi}} dE' G(r, \Delta v) Q_e (E',r)
\label{solution_transport1}
\end{equation}
with the boundary condition $\frac{dn}{dE} (r_h) = 0$, where
$r_h$ is the radius of the diffusion zone. For Draco, $r_h$ is taken to be 2.5 $kpc$ \cite{Colafrancesco:2006he,McDaniel:2017ppt}. 
$G(r, \Delta v)$ is the Green's function of the equation 
and has a form

\begin{eqnarray} 
G(r, \Delta v) &=& \frac{1}{\sqrt{4 \pi \Delta v}} \sum_{n = -\infty}^{n = \infty}
(-1)^n \int_0^{r_h} dr' \frac{r'}{r_n} 
\left( \frac{\rho_{\chi}(r')}{\rho_{\chi}(r)} \right)^2 \nonumber \\
&&\left[ \exp\left( -\frac{(r' - r_n)^2}{4 \Delta v}\right) - 
\exp\left( -\frac{(r' + r_n)^2}{4 \Delta v}\right) \right]\nonumber\\
\label{Greens}
\end{eqnarray}
with $r_n = (-1)^n r + 2 n r_h$. $\sqrt{\Delta v}$ is
called the diffusion length scale, expressed as

\begin{equation} 
\Delta v = \int_E^{E'} d\widetilde{E} \frac{D(\widetilde{E})}{b(\widetilde{E})}
\label{v_E}
\end{equation}
$\sqrt{\Delta v}$ determines the distance 
traveled by an electron as it loses energy from $E'$ to $E$. For small galaxies like dSphs, the mean value of 
this length scale is expected
to be larger than $r_h$ even for non-conservative choices of $D_0$ and $B$. We shall discuss this in detail in section \ref{astrophysics effect}.
After interacting with the magnetic field $B$
present inside the galaxy, the produced electron/positron distribution 
$\frac{dn}{dE}$ will
emit synchrotron radiation (with frequency $\nu$)  
at a rate governed by the synchrotron emission
power $P_{Synch} (\nu, E, B)$ \cite{Colafrancesco:2005ji,Beck:2015rna,Colafrancesco:2006he,McDaniel:2017ppt,Storm:2016bfw}.

\begin{figure}[ht!]
\centering
\includegraphics[height=0.35\textwidth, angle=0]{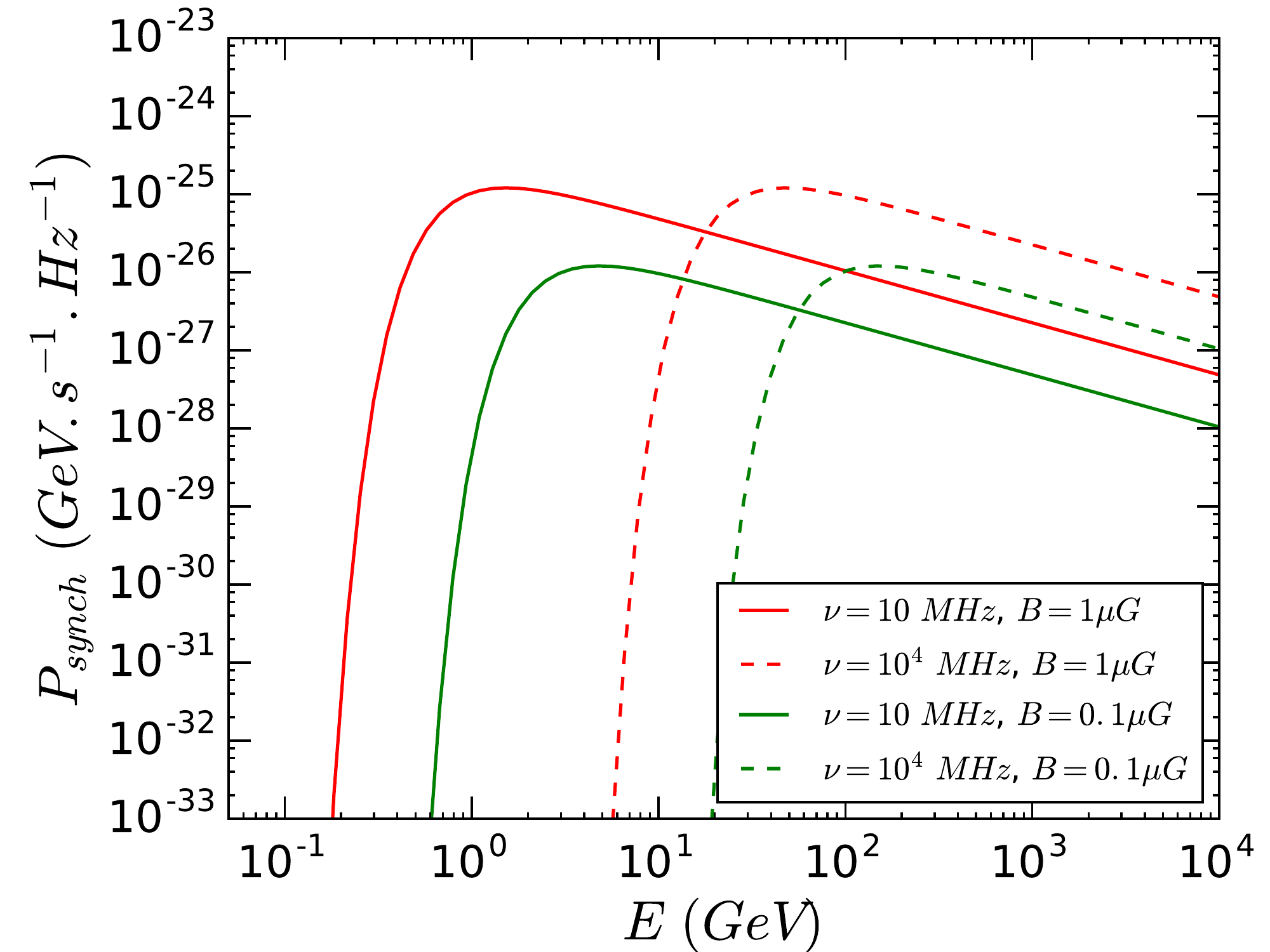}
\caption{Synchrotron power spectrum ($P_{Synch}$) vs. energy ($E$) 
at two frequencies; $10$ $MHz$ (solid lines) and $10^4$ $MHz$ (dashed lines), with 
two different magnetic fields; $B = 1$ $\mu G$ (red) and 0.1 $\mu G$ (green).}
\label{P_E}
\end{figure}

In Figure \ref{P_E}, we have shown the dependence of the synchrotron emission
power on the electron energy $E$ for two different magnetic fields 
($B =$ 1 and 0.1 $\mu G$)
and for two frequencies ($\nu =$ $10$ and $10^4$ $MHz$). For each frequency
shown, it is clear that a stronger magnetic field always intensifies the power spectrum.

The synchrotron emissivity $J_{Synch}$ as a function of frequency $\nu$ and radius $r$
is obtained by folding $\frac{dn}{dE} (r, E)$ with $P_{Synch} (\nu, E, B)$ as

\begin{equation} 
J_{Synch} (\nu, r) = 2 \int_{m_e}^{m_{\chi}} dE \frac{dn}{dE} (r, E) P_{Synch} (\nu, E, B)
\label{j_Synch}
\end{equation}
Finally, the approximate radio synchrotron flux can be obtained by
integrating $J_{Synch}$ over the diffusion size ($r_h$) of the
dSph \cite{Colafrancesco:2005ji,Beck:2015rna,Colafrancesco:2006he,McDaniel:2017ppt,Storm:2016bfw}

\begin{equation} 
S_{\nu} (\nu) = \frac{1}{L^2} \int dr r^2 J_{Synch} (\nu, r)
\label{S_nu}
\end{equation}
where $L$ is the luminosity distance of the dSph. For Draco, $L \sim 80$ $kpc$ \cite{McDaniel:2017ppt,Geringer-Sameth:2014yza}).

\section{Effect of various astrophysical parameters}\label{astrophysics effect}

Let us now take a closer look at the astrophysical effects encapsulated in 
equations \ref{transport1} and \ref{energy_loss}. Such effects are driven by the
diffusion coefficient $D_0$ and the electromagnetic energy loss coefficient
$b(E)$, which determine the steady state
$e^{\pm}$ distribution $\frac{dn}{dE} (r, E)$ for a given $\frac{dN^e}{dE}$. 
As has been already mentioned,
$b(E)$ is implicitly dependent on the galactic magnetic field $B$. 

We use again the two DM masses ($m_{\chi} = 300$ $GeV$ and 5 $TeV$) for studying the effects of $D_0$ and $b(E)$.
To have some idea on these effects separately, as well as their contribution
in an entangled fashion, we have considered the implication of equation
\ref{transport1} for three different scenarios:

\begin{itemize}
\item NSD: considering only the effect of energy loss term $b(E)$ and neglecting the spatial diffusion $D(E)$ (i.e. solution of equation
\ref{transport1}
by setting $D(E) = 0$).

\item Nb: considering the effect of diffusion parameter
$D_0$ and neglecting the energy loss term $b(E)$ (i.e. solution of equation \ref{transport1} by setting $b(E) = 0$).

\item SD+b: considering the effects of
both the diffusion parameter $D_0$ and the energy loss term $b(E)$ (i.e. the complete
solution as explained in equation \ref{solution_transport1}).
\end{itemize}

\begin{figure}[ht!]
\centering
  \includegraphics[height=0.35\textwidth, angle=0]{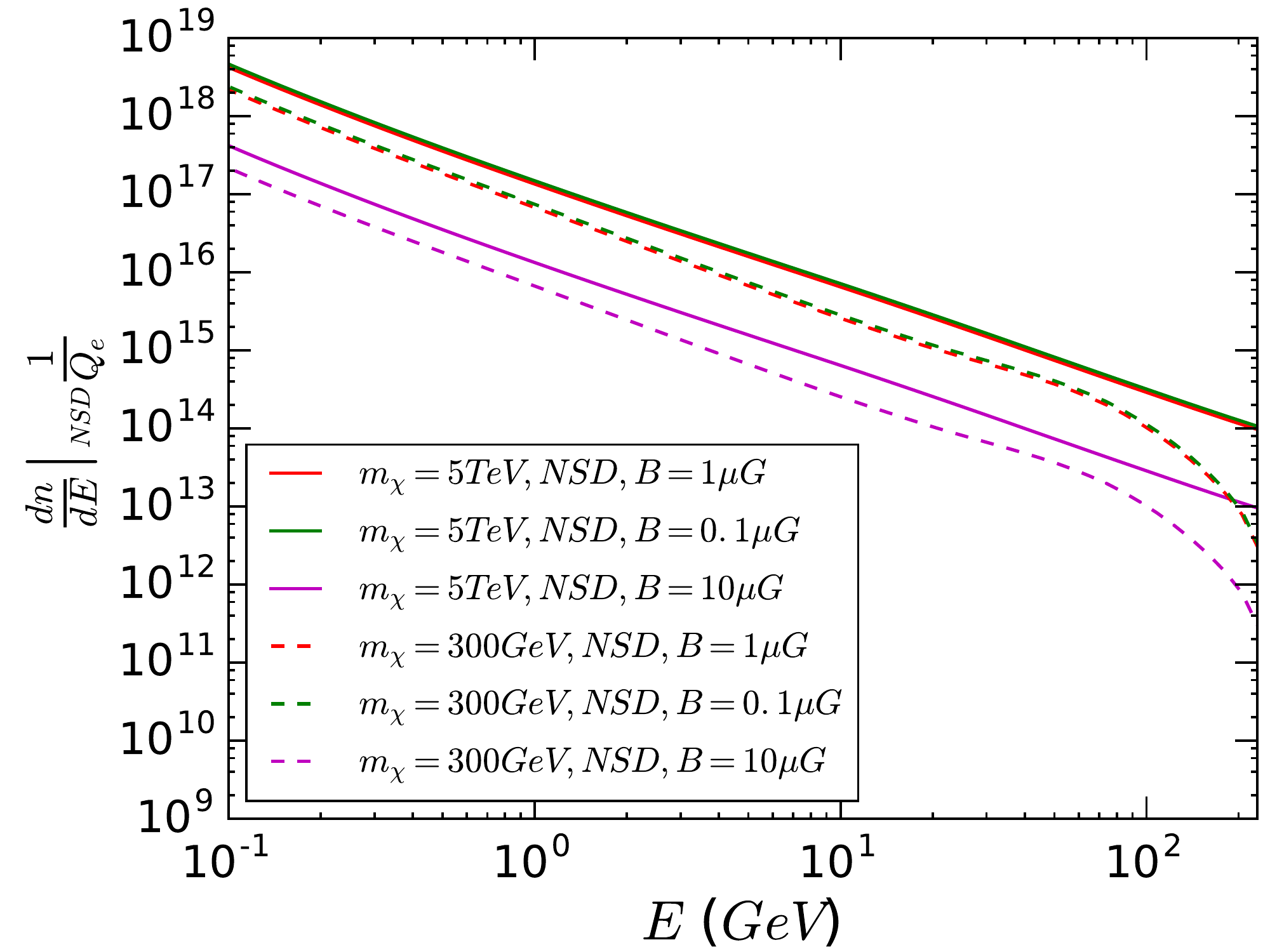}
  \caption{$\frac{dn}{dE} \frac{1}{Q_e}$ vs. electron energy $E$ plot
  for two DM masses, 5 $TeV$ (solid lines) and 300 $GeV$ (dashed lines)
  with magnetic fields $B$ = 1 $\mu G$ (red), 0.1 $\mu G$ (green) and 10 $\mu G$ 
  (magenta) in the scenario
  where diffusion in the system has been neglected (NSD). Annihilation 
  channel is $b\bar{b}$ with annihilation rate
  $\langle \sigma v\rangle$ = $10^{-26} \mbox{cm}^3 \mbox{s}^{-1}$.}
\label{NSD}  
\end{figure}

For the NSD scenario, solution of equation \ref{transport1} 
has a simpler form \cite{Colafrancesco:2005ji}:

\begin{equation} 
\left.\frac{dn}{dE} (r, E)\right|_{\mbox{NSD}} = \frac{1}{b(E)} \int_E^{m_{\chi}} dE' Q_e (E', r)
\label{solution_transport2}
\end{equation} 
Note that, $b(E)$ increases with $E$ (from equation \ref{energy_loss}). Its effect on the number density in different 
energy regions is more prominent for this NSD scenario where we neglect the effect of distribution. 
The number density decreases in the high energy region due to the combined effect of the 
$\frac{1}{b(E)}$ suppression as well as the lower 
limit of the integration in equation \ref{solution_transport2}. This can also be 
seen from Figure \ref{NSD} where we have plotted the
ratio of $\frac{dn}{dE}$ and the 
source function
$Q_e$ (which determines the initial electron flux due to DM annihilation) against $E$.
The term $\frac{dn}{dE} \frac{1}{Q_e}$ (in unit of $s^{-1}$) 
essentially determines how the shape of the
initial distribution $Q_e$ gets
modified due to the effect of various astrophysical processes.   
As discussed in section \ref{Essential Processes}, 
the energy loss term $b(E)$ depends on the square
of the magnetic field $B$ through synchrotron loss. Thus $b(E)$ increases with 
increase of $B$, which in turn will reduces the $\frac{dn}{dE}$ at all $E$. 
This phenomenon can also be observed in Figure \ref{NSD} where the red and magenta lines 
indicate the distributions for $B = 1$ and 10 $\mu G$\footnote{$B = 10$ $\mu G$ has been used
in this Figure for the sake of comparison with 1 $\mu G$, just to see the effect of 
`large $B$'. In our prediction on observable 
radio flux, more conservative (and perhaps realistic) values of $B$ have been used.}, respectively. 
For comparison, we have also shown the cases for $B = 0.1$ $\mu G$ (green lines). Note that, the later case coincides with the case for $B = 1$ $\mu G$. This is due to the fact that the 
magnetic field dependence (through synchrotron loss) in the energy loss term $b(E)$ gets suppressed by the Inverse Compton ($b_{IC}$) term for lower $B$ ($B < 1$ $\mu G$), which can be clearly seen from equation \ref{energy_loss}.  
Also, for heavier DM masses the energetic electrons are produced in greater 
abundance (as seen from Figure \ref{Q_E}) which can lead to a larger 
$\frac{dn}{dE} \frac{1}{Q_e}$. The effects of two different DM masses, $m_{\chi} = 300$
$GeV$ and 5 $TeV$, have been shown in this figure by dashed and solid lines respectively.

\begin{figure*}[ht!]
\centering
  \includegraphics[height=0.33\textwidth, angle=0]{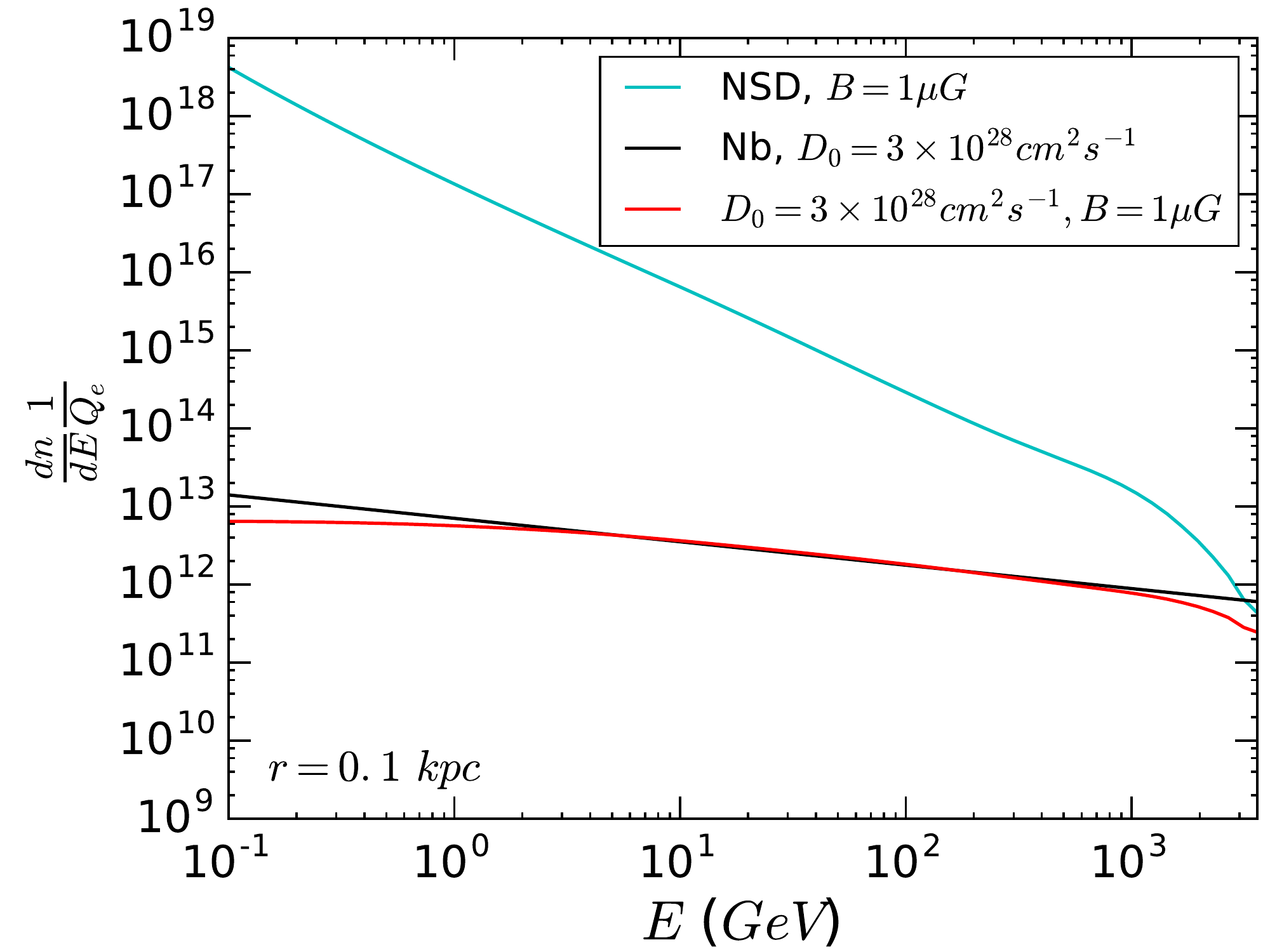}
  \hspace{2mm}
  \includegraphics[height=0.33\textwidth, angle=0]{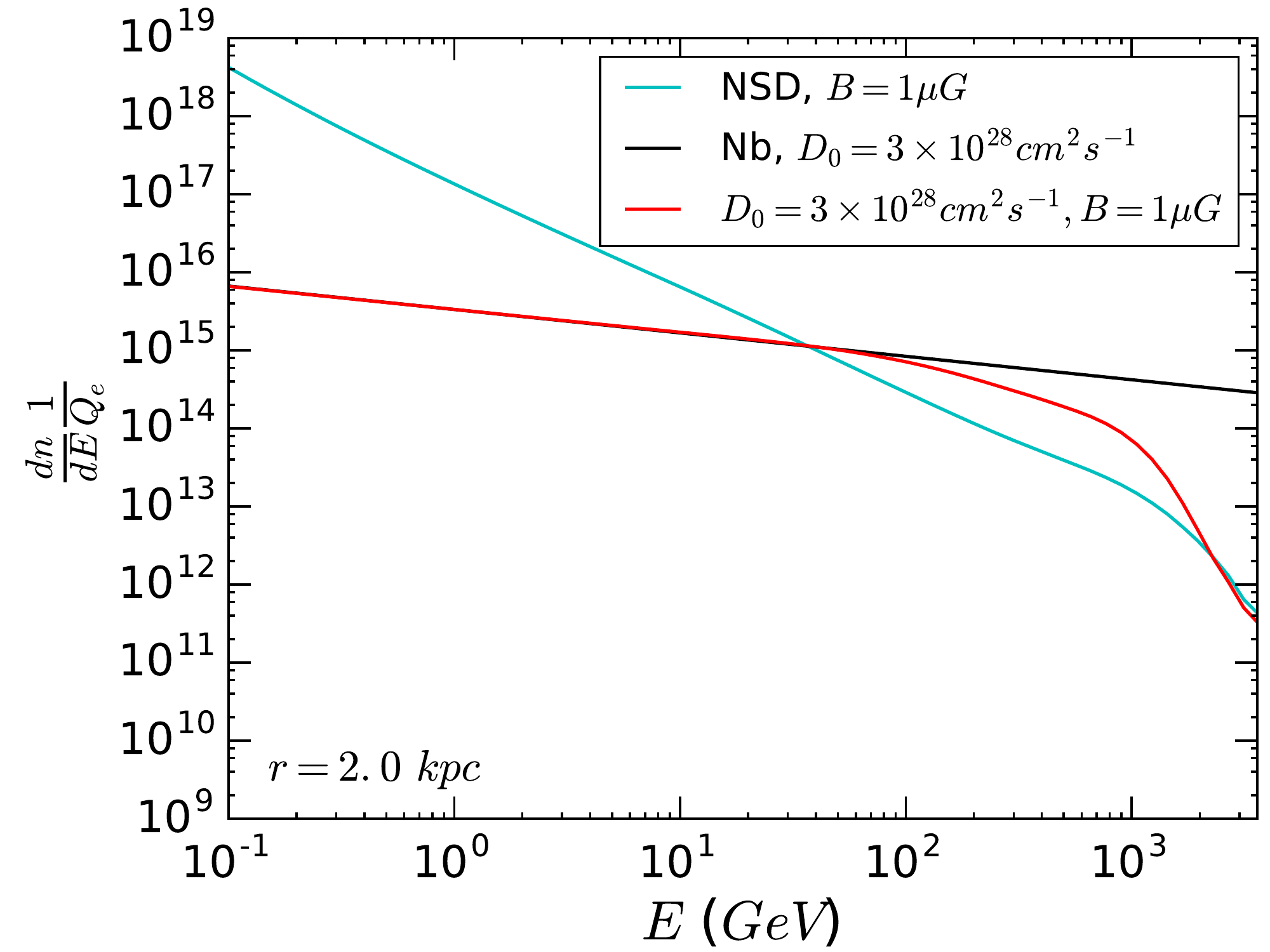}
  \caption{$\frac{dn}{dE} \frac{1}{Q_e}$ vs. electron energy $E$
  at two different radii, $r = 0.1 kpc$ ({\it left panel}) and $r = 2.0 kpc$ ({\it right panel})
   for $m_{\chi} = 5$ $TeV$ in three scenarios, 
  NSD, Nb and SD+b. Cases including diffusion (Nb and SD+b)
  have $D_0 = 3 \times 10^{28} \mbox{cm}^2 \mbox{s}^{-1}$ and
  cases including energy loss effect (NSD and SD+b)
  have magnetic field $B$ = 1 $\mu G$. For all cases annihilation 
  channel is $b\bar{b}$ with annihilation rate
  $\langle \sigma v\rangle$ = $10^{-26} \mbox{cm}^3 \mbox{s}^{-1}$.}
\label{NSD_total}  
\end{figure*}

Now, for the Nb scenario, the solution becomes

\begin{equation} 
\left.\frac{dn}{dE} (r, E)\right|_{\mbox{Nb}} = \frac{1}{D(E)} f(r, r_h) Q_e(r,E)
\label{solution_transport3}
\end{equation}
with the same boundary condition as what
we assumed for equation \ref{solution_transport1}. Here

\begin{equation} 
f(r, r_h) = \int_{r' = r}^{r_h} dr' (\frac{1}{r'^2}) 
\left\lbrace \int_{\widetilde{r} = 0}^{r'} d\widetilde{r} 
\hspace{1mm} \widetilde{r}^2
\left( \frac{\rho_{\chi}(\widetilde{r})}{\rho_{\chi}(r)} \right)^2 \right\rbrace   
\end{equation}
Since the effect of $b(E)$ is absent
in this case, the distribution $\frac{dn}{dE}$ will 
not depend on the magnetic field $B$. Figure \ref{NSD_total} shows the comparison
between the NSD case (cyan lines)
and the Nb case (black lines) at two different radii
($r = 0.1$ $kpc$; left panel and $r = 2$ $kpc$; right panel)
for $B = 1$ $\mu G$ and $D_0 = 3 \times 10^{28} \mbox{cm}^2 \mbox{s}^{-1}$.
It is clear from equation \ref{solution_transport3} that the ratio 
$\frac{dn}{dE} \frac{1}{Q_e}$ for the Nb scenario
will not depend on the initial spectrum $Q_e$ and its energy profile
will follow the energy dependence of $\frac{1}{D(E)}$.
This presence of $D(E)$ in the denominator leads to 
$\frac{dn}{dE} \frac{1}{Q_e} \sim E^{-0.3}$ in the limit $b(E) \to 0$.
This can also be verified from the relatively flat curves for Nb cases in this figure.

\begin{figure*}[ht!]
\centering
\includegraphics[height=0.33\textwidth, angle=0]{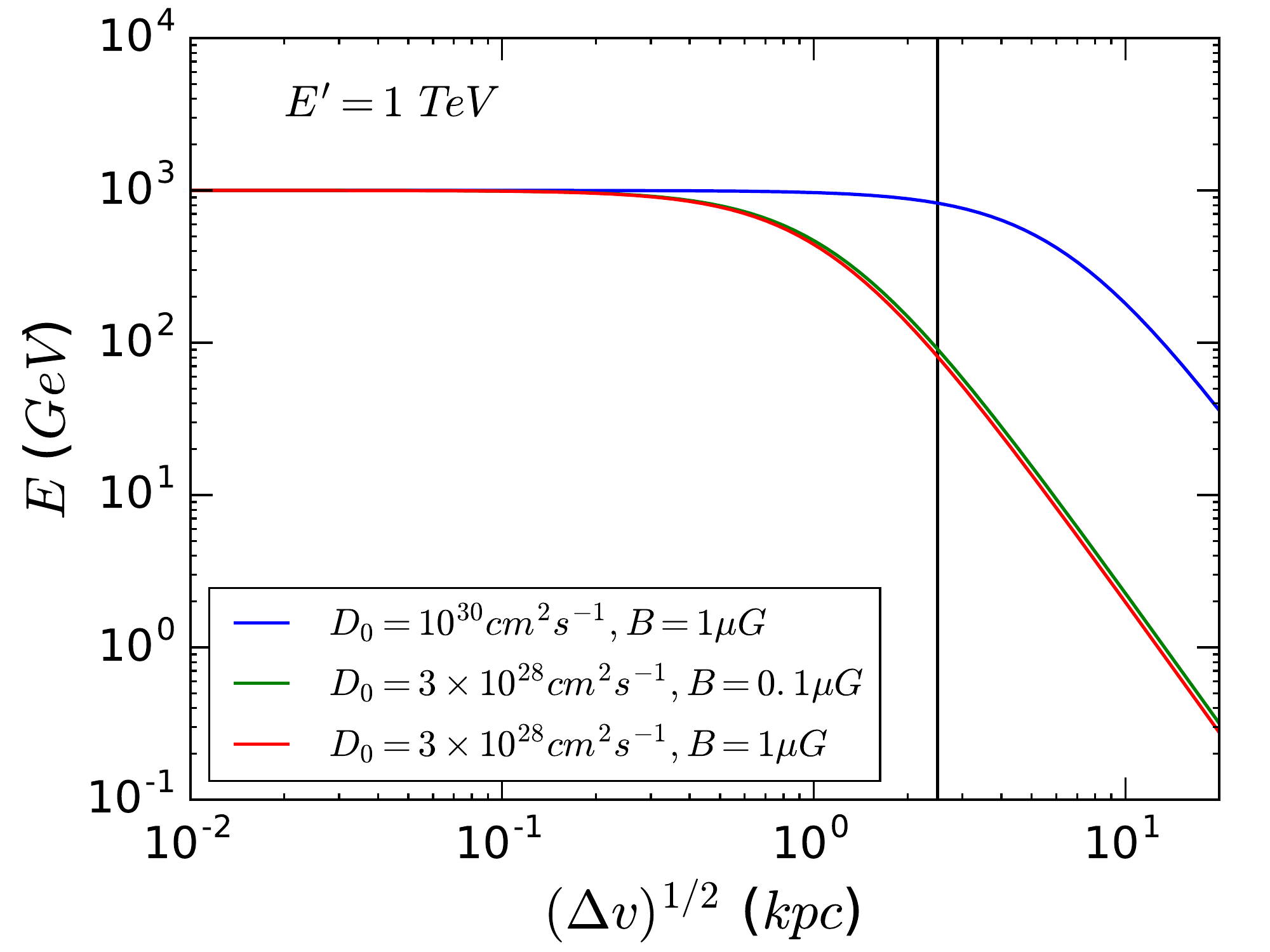}
\hspace{2mm}
\includegraphics[height=0.33\textwidth, angle=0]{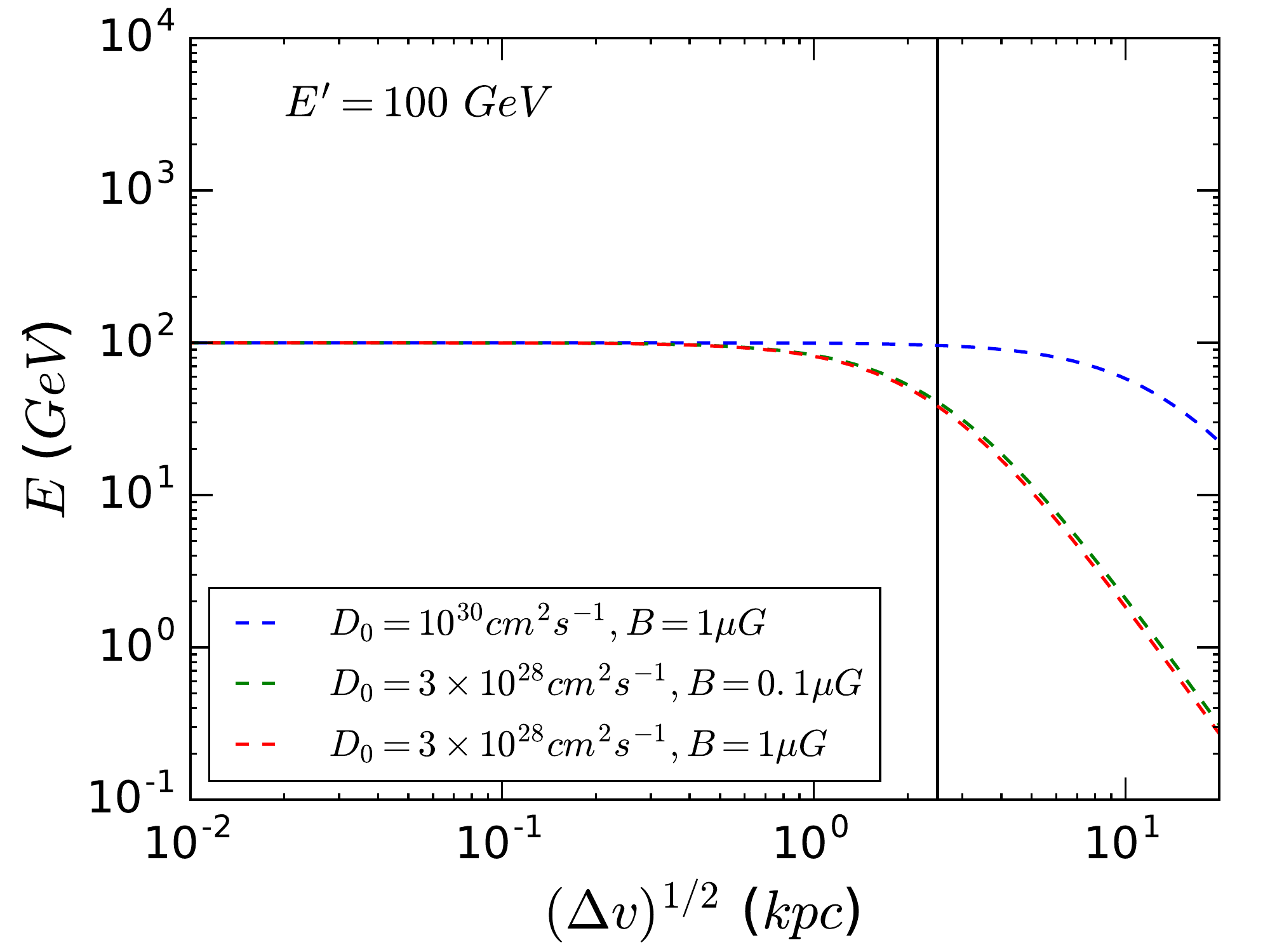}
\caption{Electron interaction energy $E$ ($GeV$) vs. diffusion length scale 
$\sqrt{\Delta v}$ (kpc)  for two different source energies, $E' = 1$ $TeV$
({\it left panel}) and $E' = 100$ $GeV$ ({\it right panel}). The vertical 
black solid lines indicate
the diffusion size of the galaxy ($r_h = 2.5$ $kpc$). Here three different
sets of astrophysical parameters have been chosen, 
$D_0 = 3 \times 10^{28} \mbox{cm}^2 \mbox{s}^{-1}$, $B = 1$ $\mu G$ (red);
$D_0 = 10^{30} \mbox{cm}^2 \mbox{s}^{-1}$, $B = 1$ $\mu G$ (blue) and
$D_0 = 3 \times 10^{28} \mbox{cm}^2 \mbox{s}^{-1}$, $B = 0.1$ $\mu G$ (green).}
\label{length_scale}  
\end{figure*}

We should mention here that neglecting diffusion is not a 
bad assumption where the length scale ($\sqrt{\Delta v}$) over which the $e^{\pm}$ losses energy is much shorter than the typical size of the system \cite{Colafrancesco:2005ji,McDaniel:2017ppt}. 
However, for smaller systems 
like dSphs the effect of diffusion cannot be neglected. This 
can be justified from Figure \ref{NSD_total} where we have plotted the SD+b 
case (i.e. taking diffusion into account along with energy
loss effect; shown by red curves) along with the two previously mentioned scenarios. 
It is clear that the NSD scenario is strikingly different from
the one which takes diffusion into account.
Addition of diffusion in the system essentially suppresses the $e^{\pm}$ distribution,
especially in the low energy energy region.
As a result, the plots for the SD+b case closely match the Nb one at lower energy,
but diverges following the NSD scenario at higher energies
where the effect of $b(E)$ dominates.
This phenomenon can also be 
explained in terms of the diffusion length scale ($\sqrt{\Delta v}$) which
is a result of combined effects of diffusion and energy loss (equation \ref{v_E}). 
It is indicative of the length covered by an electron as it loses energy from the source energy $E'$ to the interaction energy $E$ which is 
typically larger than the size of the dSph.
In order to determine the minimum energy that an electron can possess before escaping
the diffusion radius, we have plotted the interaction energy $E$ vs. $\sqrt{\Delta v}$
in Figure \ref{length_scale} for different sets of $D_0$ and $B$. The left and right panels
correspond to the source energy $E' = 1$ $TeV$ and 100 $GeV$ respectively.
The diffusion radius (for Draco) $r_h = 2.5$ $kpc$ is shown by the black solid lines
for both cases. The minimum energy is thus calculated from the point where the model crosses this line.
A higher $D_0$ 
will increase this minimum value of interaction energy and as a result, it reduces the $e^{\pm}$ number density by making them leave the diffusion zone earlier.  
Note that, the cases with $B = 0.1$ $\mu G$ practically coincides with that for $B = 1$ $\mu G$. This is due to the fact that $b(E)$ remains unchanged for $B \leq 1$ $\mu G$, as already been mentioned.
On the other hand, we have explicitly checked that, a larger value of $B$ (say 
$B = 10$ $\mu G$; not shown in the plot) will help the electron in losing energy more quickly before escaping the diffusion zone, causing a suppression in the number density of high energy electrons and an enhancement of the low energy electrons.

\begin{figure*}[ht!]
\centering
  \includegraphics[height=0.33\textwidth, angle=0]{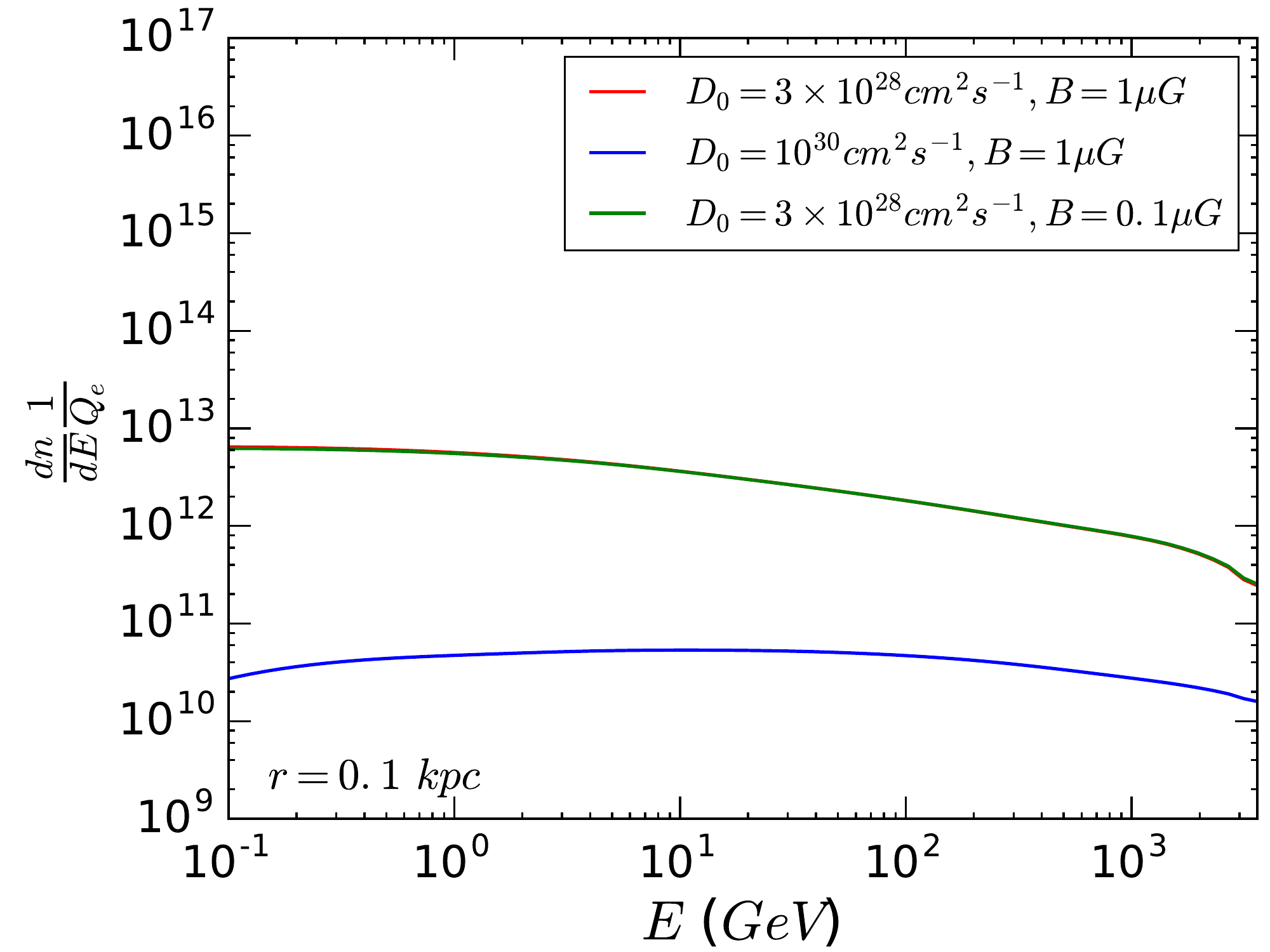}
  \hspace{2mm}
  \includegraphics[height=0.33\textwidth, angle=0]{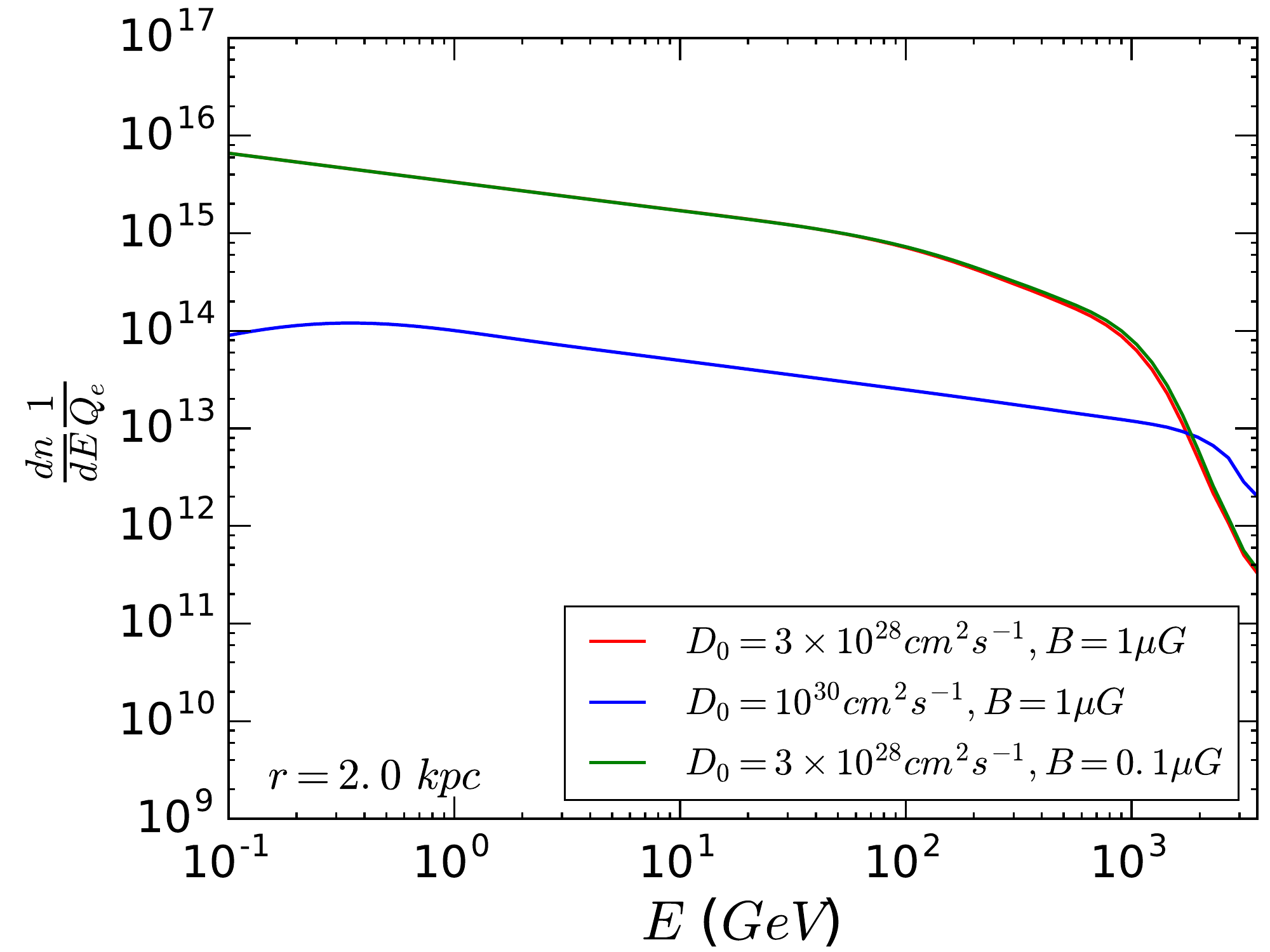}
  \includegraphics[height=0.33\textwidth, angle=0]{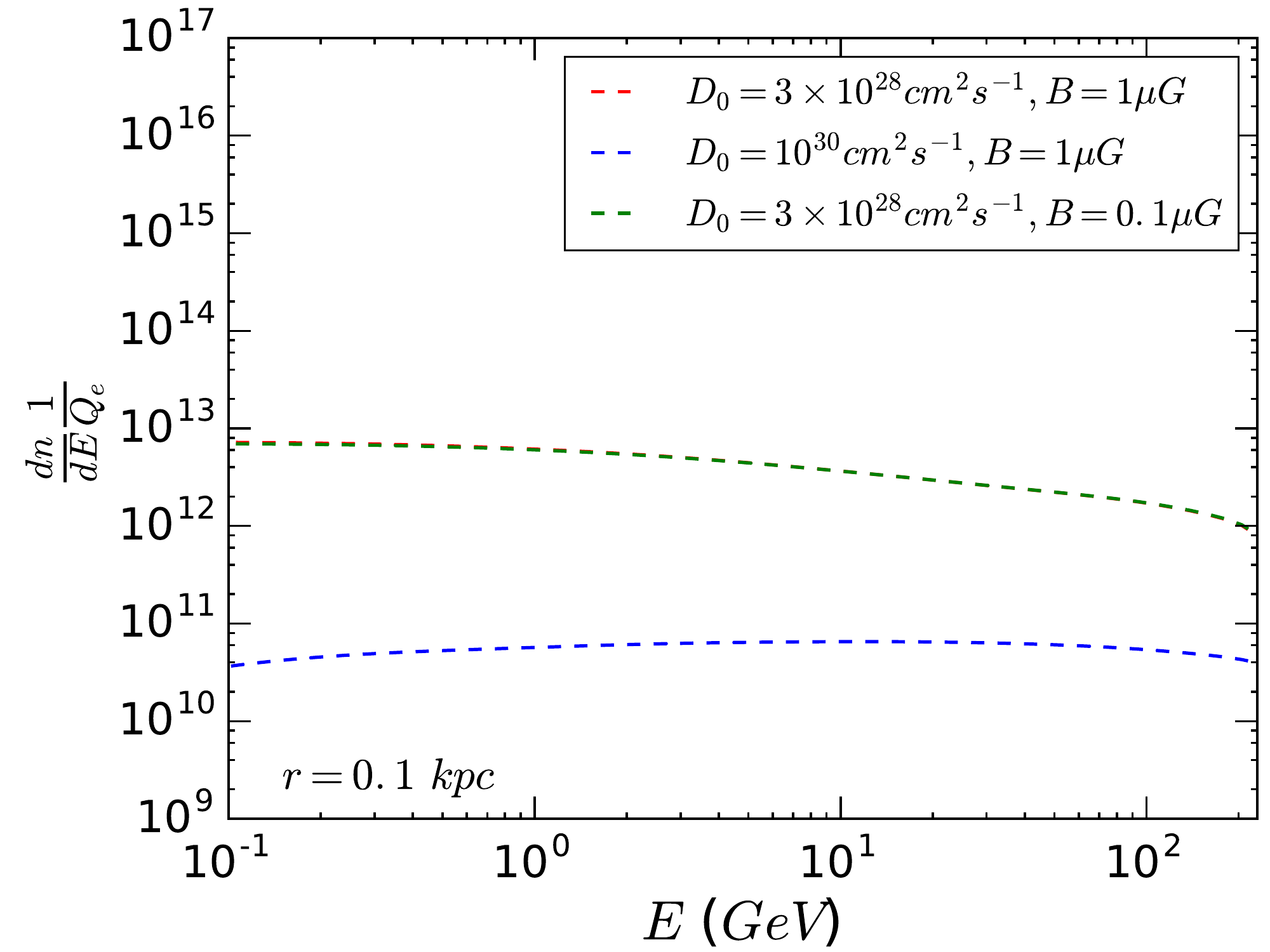}
  \hspace{2mm}
  \includegraphics[height=0.33\textwidth, angle=0]{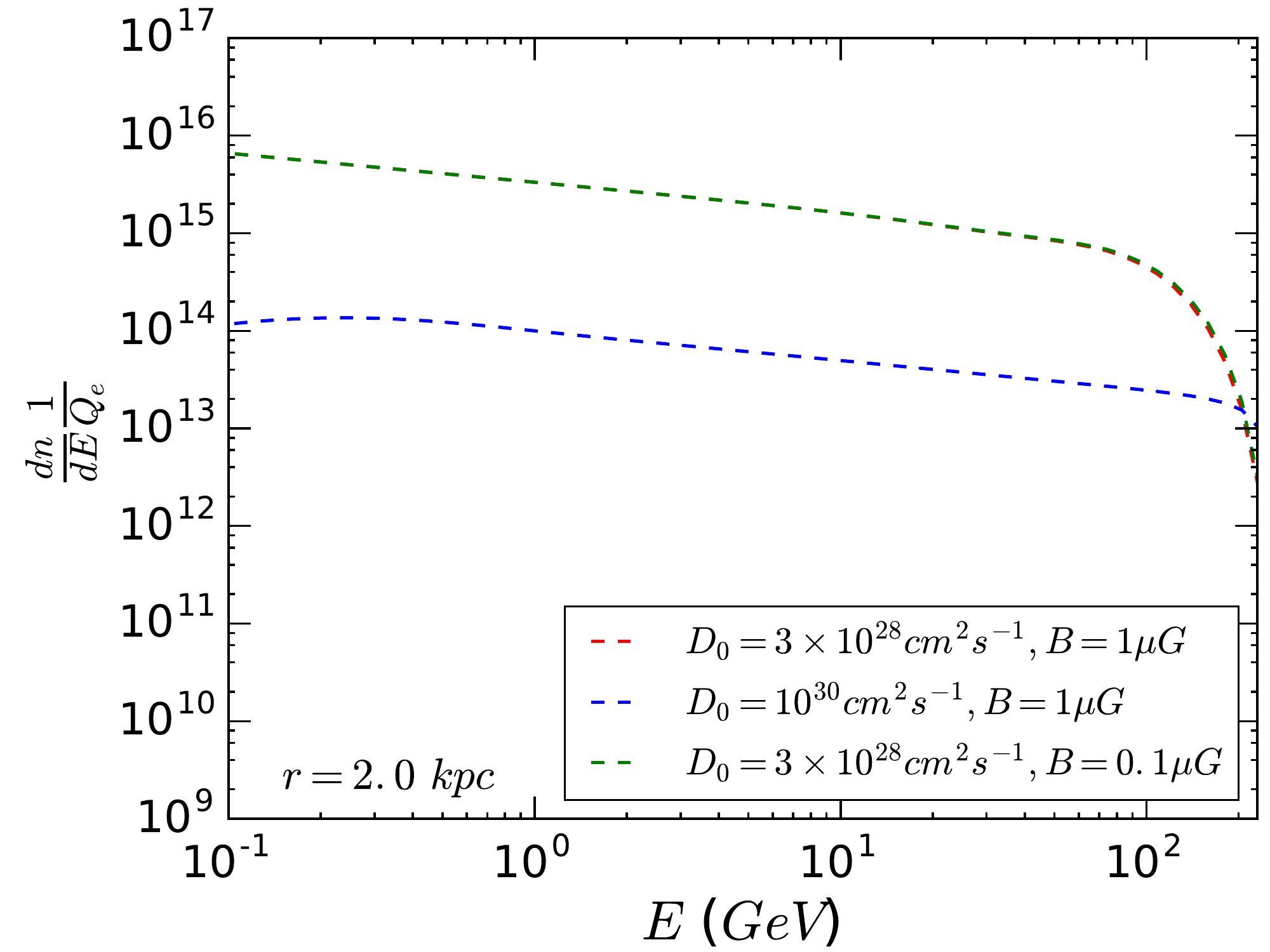}
  \caption{$\frac{dn}{dE} \frac{1}{Q_e}$ vs. electron energy $E$
  at two different radii, $r = 0.1 kpc$ ({\it left panel}) and $r = 2.0 kpc$ ({\it right panel})
  for two DM masses, 5 $TeV$ ({\it upper panels} -- solid lines) and 300 $GeV$ ({\it lower panels} -- dashed lines) in
  the SD+b scenario.
  Astrophysical parameters considered here have been mentioned in the corresponding legends.
  For all cases annihilation 
  channel is $b\bar{b}$ with annihilation rate
  $\langle \sigma v\rangle$ = $10^{-26} \mbox{cm}^3 \mbox{s}^{-1}$.}
  \label{total}
\end{figure*}

\begin{figure*}[ht!]
\centering
  \includegraphics[height=0.33\textwidth, angle=0]{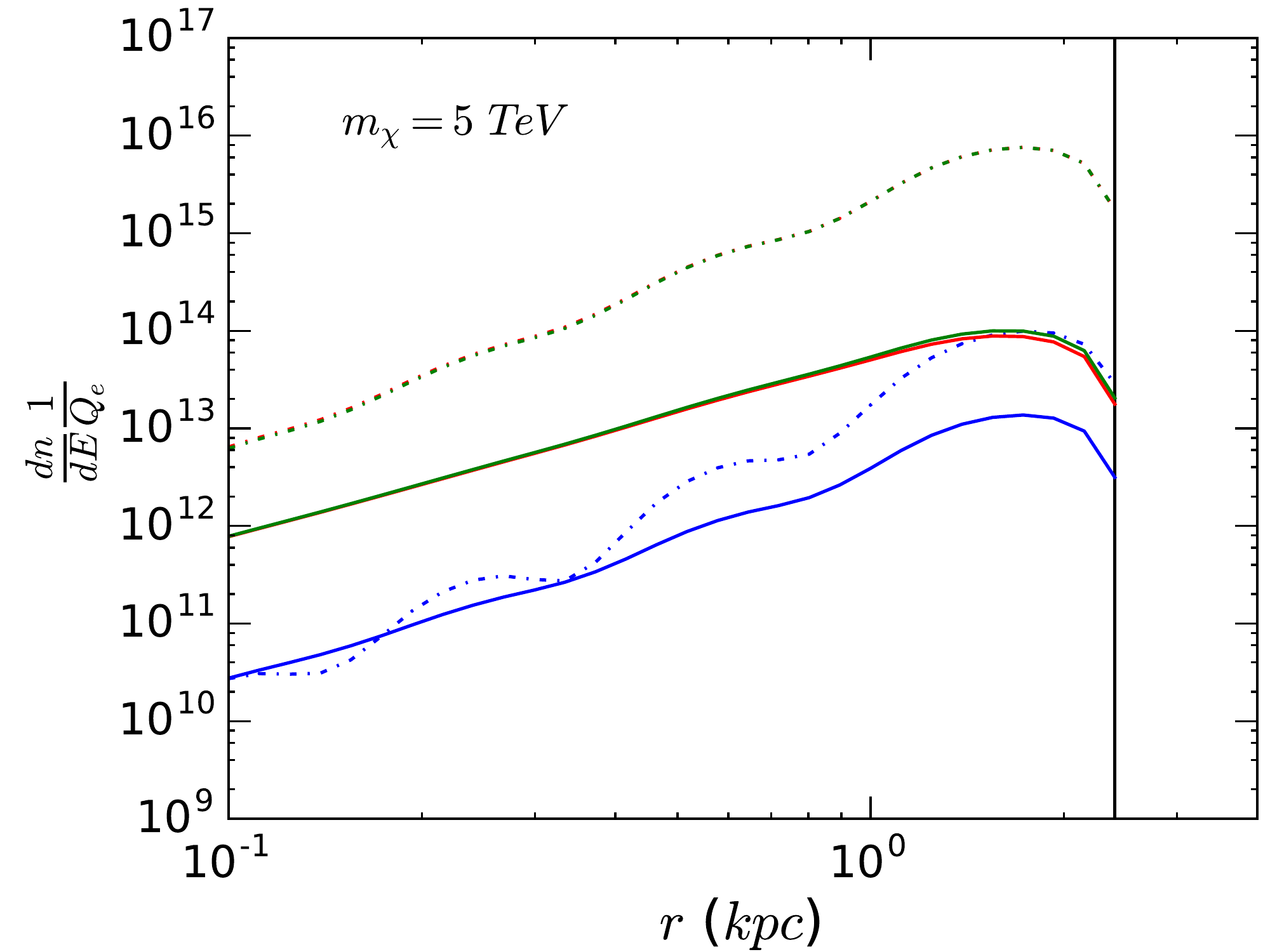}
  \hspace{2mm}
  \includegraphics[height=0.33\textwidth, angle=0]{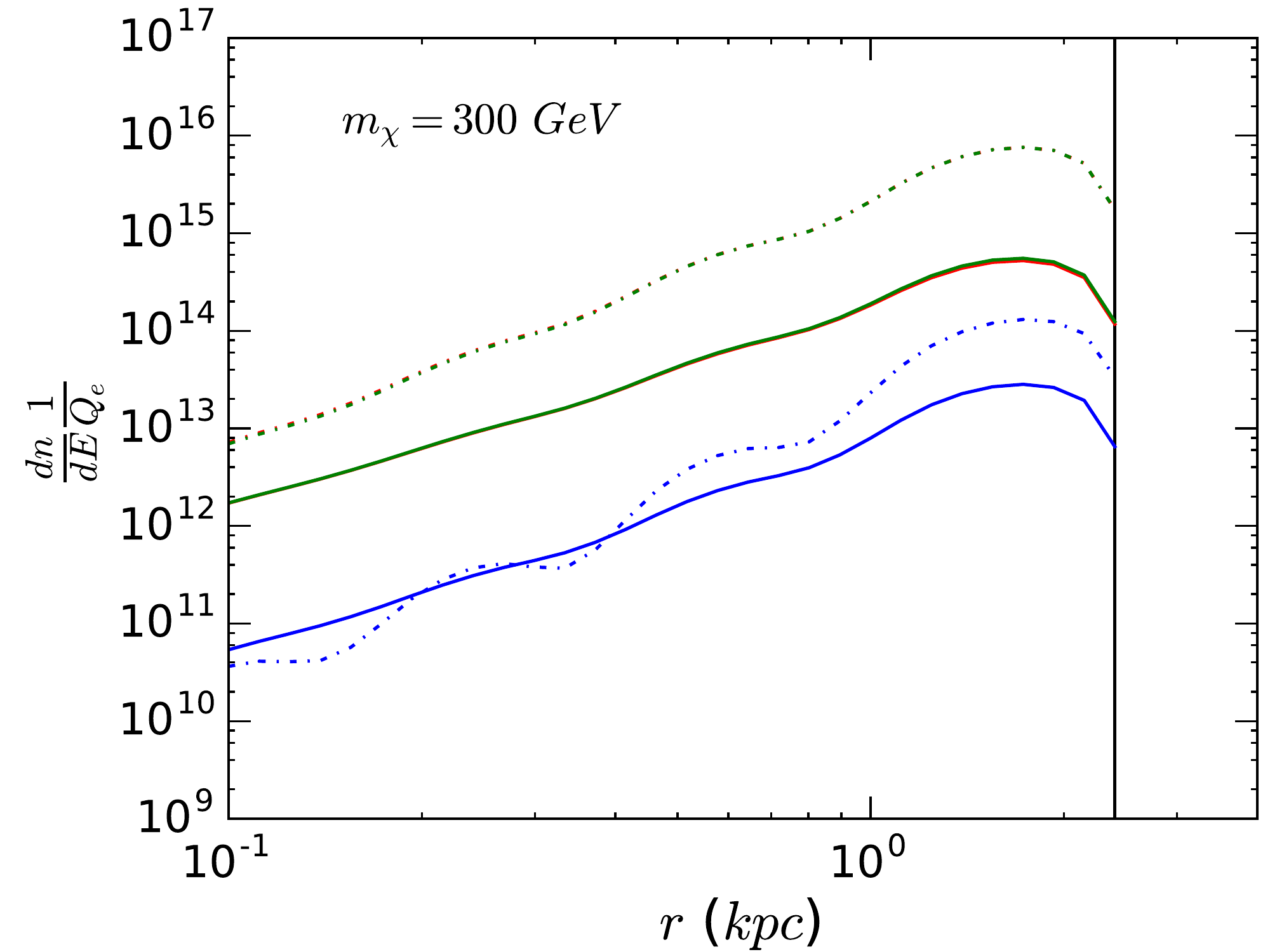}
  \caption{{\it Left panel:} $\frac{dn}{dE} \frac{1}{Q_e}$ vs. $r$ plot
  for $m_{\chi} = 5$ $TeV$ at two different electron energies, 
  $E = 0.1$ $GeV$ (dashed dotted lines) 
  and $E = 1$ $TeV$ (solid lines)
  in the SD+b scenario. The vertical black solid line indicates the diffusion size ($r_h$) of the dSph.
  Three different sets of astrophysical parameters have been considered, 
  $D_0 = 3 \times 10^{28} \mbox{cm}^2 \mbox{s}^{-1}$, $B = 1$ $\mu G$ (red);
  $D_0 = 10^{30} \mbox{cm}^2 \mbox{s}^{-1}$, $B = 1$ $\mu G$ (blue)
  and
  $D_0 = 3 \times 10^{28} \mbox{cm}^2 \mbox{s}^{-1}$, $B = 0.1$ $\mu G$ (green).
  Annihilation 
  channel is $b\bar{b}$ with annihilation rate
  $\langle \sigma v\rangle$ = $10^{-26} \mbox{cm}^3 \mbox{s}^{-1}$.
  {\it Right panel:} Same as the {\it left panel} but at $E = 0.1$ $GeV$ (dashed dotted lines) 
  and $E = 100$ $GeV$ (solid lines) for $m_{\chi} = 300$ $GeV$.}
  \label{total_r}
\end{figure*}

Similar conclusions can also be drawn from Figure \ref{total}. To illustrate
the effects of $D_0$ and $B$ in the SD+b
scenario, we have plotted $\frac{dn}{dE} \frac{1}{Q_e}$ against $E$ for the 
DM masses 5 $TeV$
(upper panels) and 300 $GeV$ (lower panels) at two different radii 0.1
$kpc$ (left panels) and 2 $kpc$ (right panels). For all cases, the dominant
annihilation channel has been assumed to be $b\bar{b}$ with $\langle \sigma v\rangle$ = $10^{-26} \mbox{cm}^3 \mbox{s}^{-1}$. Cases with
different $D_0$ and $B$ are indicated by different color coding as mentioned
in the insets of corresponding figures.
Similarly, Figure \ref{total_r} shows the variation of the same
with respect to radius $r$ for those two DM masses ($m_{\chi} = 5$ $TeV$ $-$ left panel and
$m_{\chi} = 300$ $GeV$ $-$ right panel). For each
DM mass we have assumed two different combinations of energies, one being low 
($E = 0.1$ $GeV$ for both DM masses) and the
other, high ($E = 1$ $TeV$ for $m_{\chi} = 5$ $TeV$ and 
$E = 100$ $GeV$ for $m_{\chi} = 300$ $GeV$). 
In both panels, the black line indicates the diffusion radius (2.5 $kpc$) for Draco. 
One can see that, large $D_0$
(blue curves) implies a lower density distribution compared to
the corresponding cases with smaller diffusion coefficient. 
The difference is more prominent for low energy.

\begin{figure*}[ht!]
\centering
  \includegraphics[height=0.36\textwidth, angle=0]{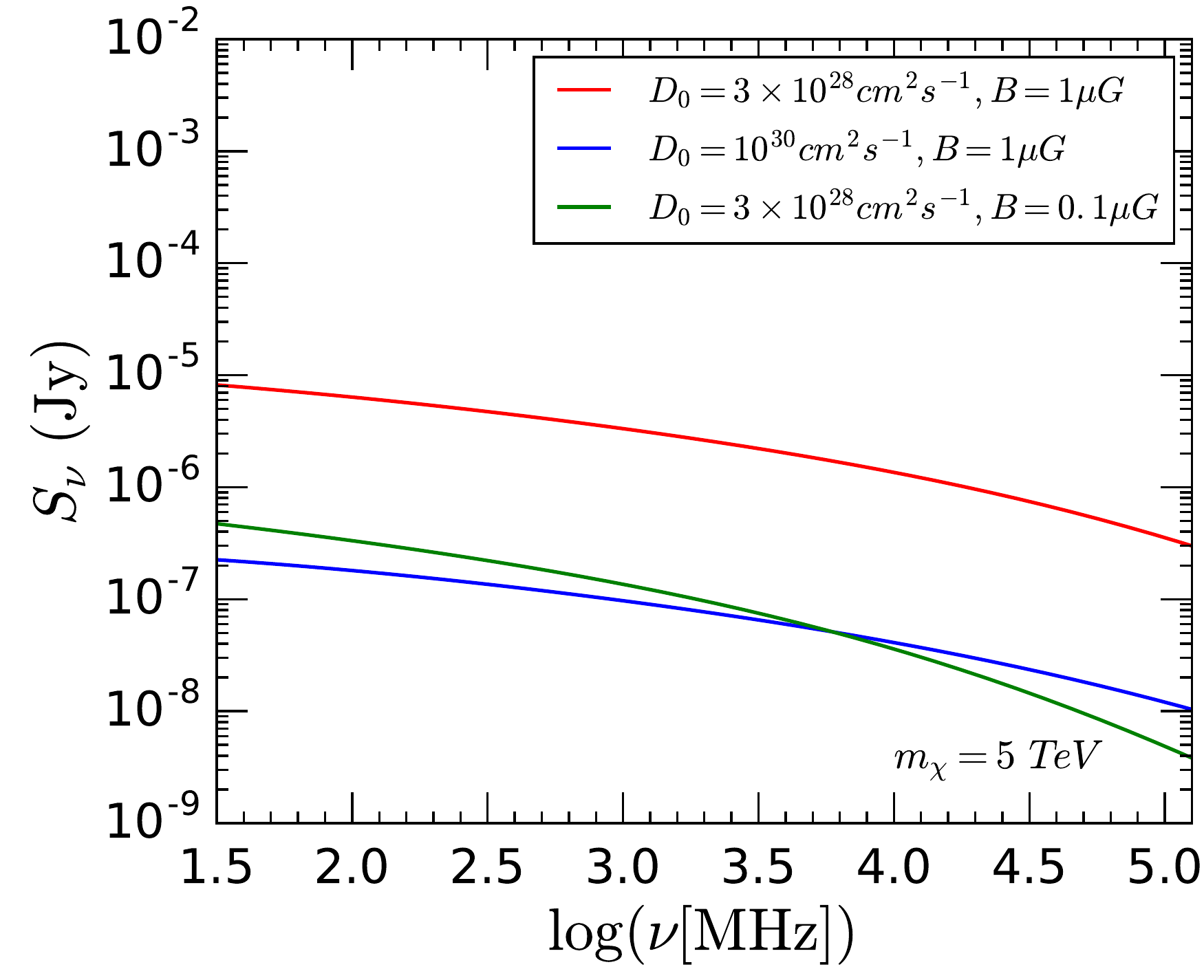}
  \hspace{2mm}
  \includegraphics[height=0.36\textwidth, angle=0]{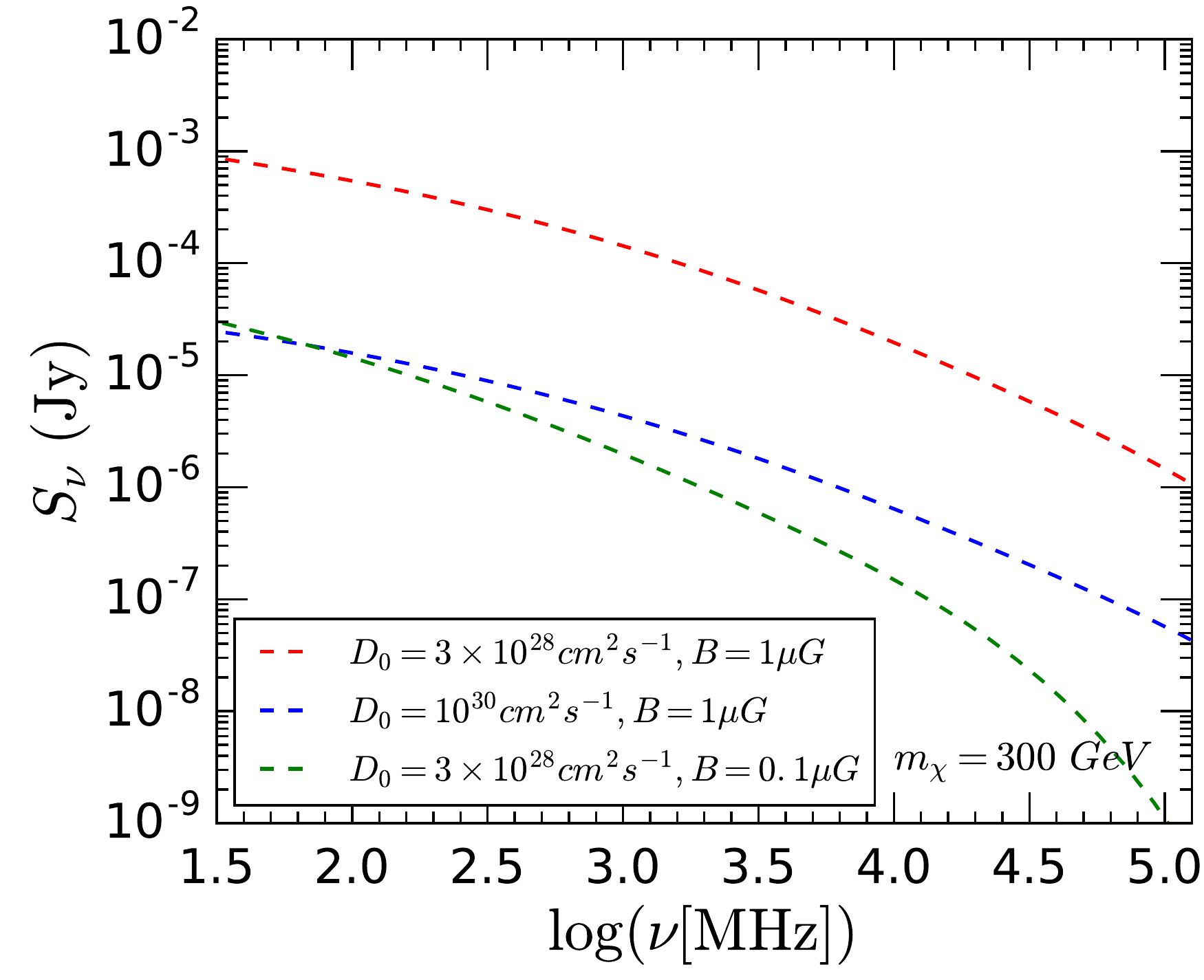}
  \caption{Radio synchrotron flux ($S_{\nu}$) vs. frequency ($\nu$) for
  two DM masses, 5 $TeV$ ({\it left panel}) and 300 $GeV$ ({\it right panel}) with
  different choices of astrophysical parameters ($D_0$ and $B$) mentioned in the legends.
  Annihilation 
  channel is $b\bar{b}$ with annihilation rate
  $\langle \sigma v\rangle$ = $10^{-26} \mbox{cm}^3 \mbox{s}^{-1}$.}
  \label{nu_S_B_D}
\end{figure*}

\begin{figure*}[ht!]
\centering
  \includegraphics[height=0.36\textwidth, angle=0]{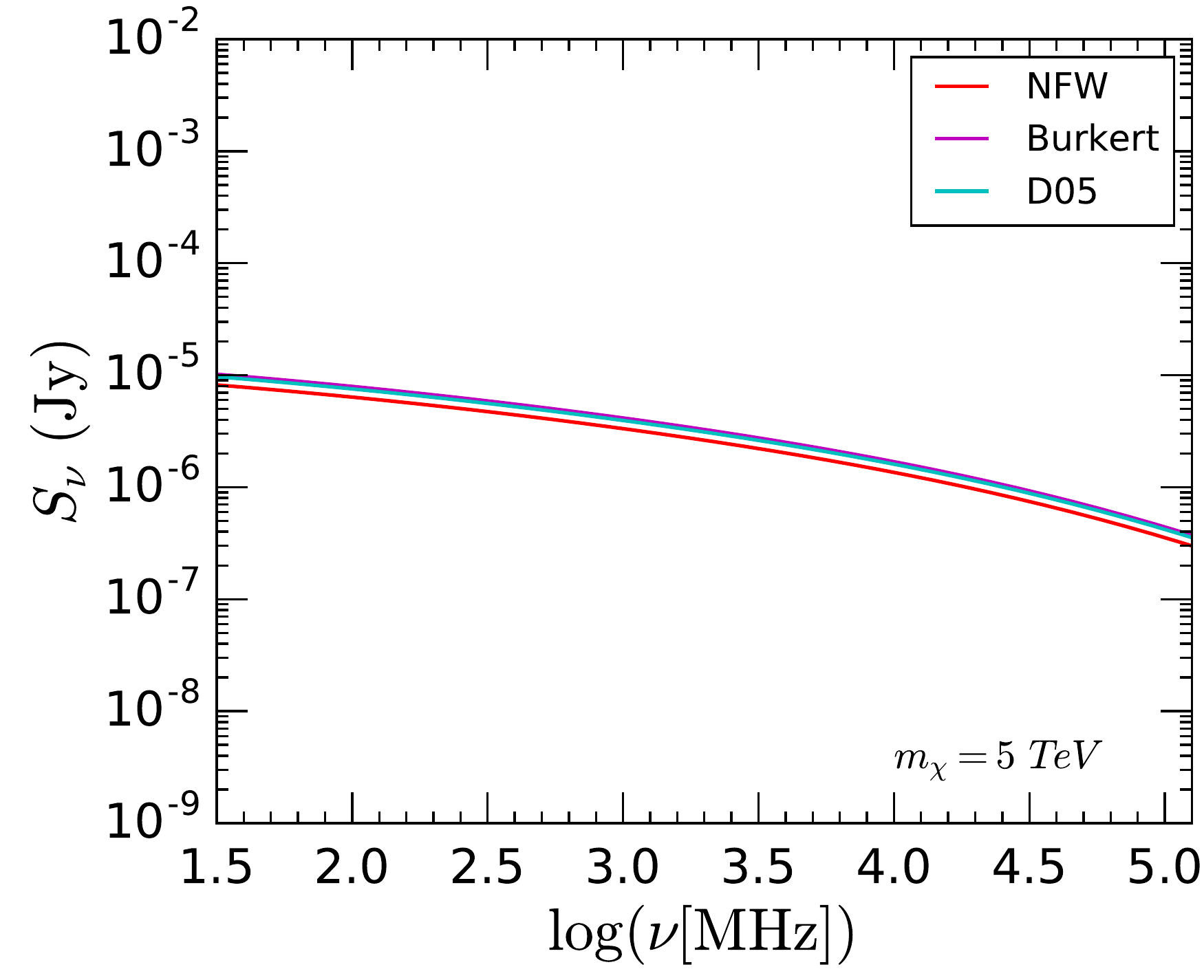}
  \hspace{2mm}
  \includegraphics[height=0.36\textwidth, angle=0]{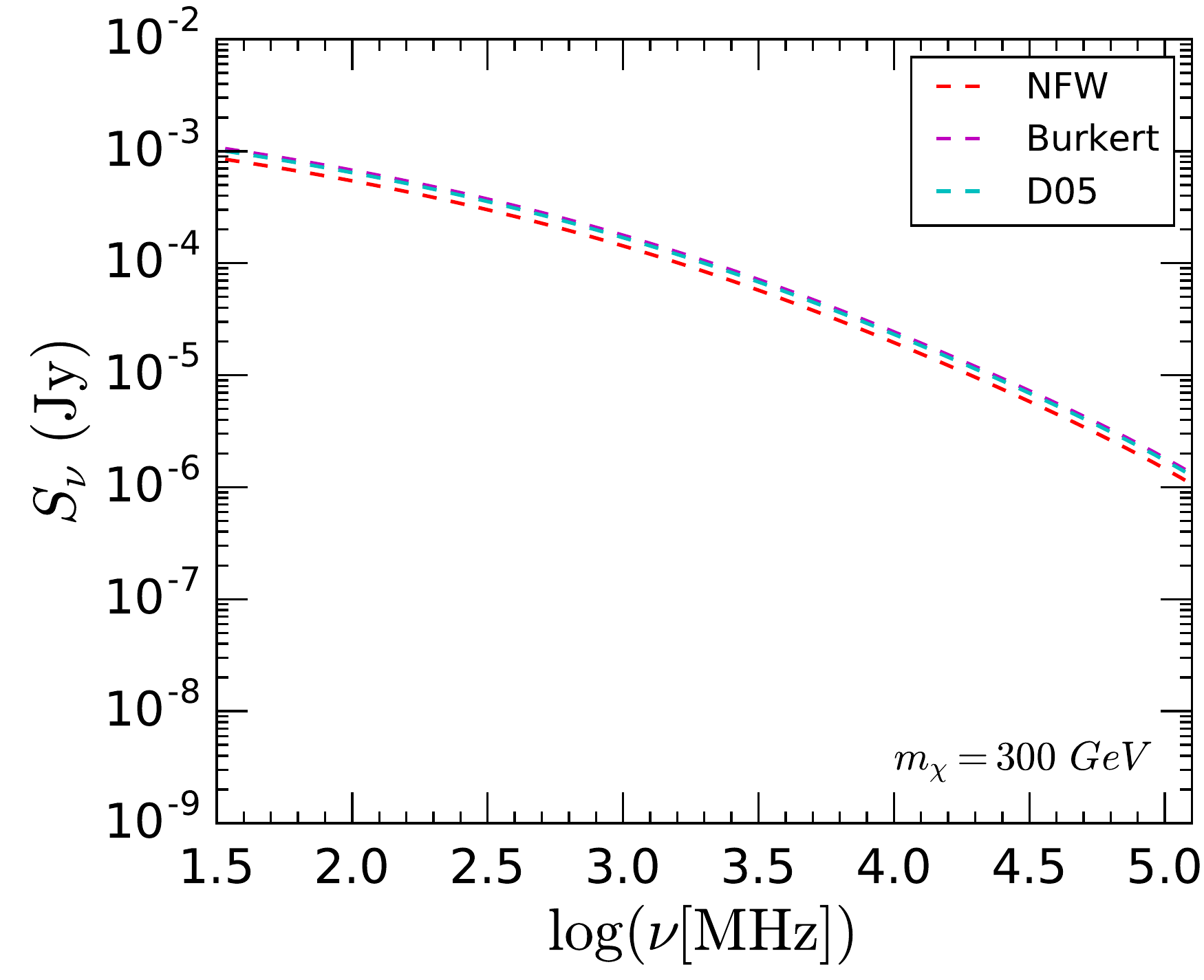}
  \caption{{\it Left panel:} Radio synchrotron flux $S_{\nu}(\nu)$ for Draco assuming 
  three different DM profiles, NFW (red), Burkert (magenta) and D05 (cyan) \cite{Colafrancesco:2006he}. DM mass
  is $m_{\chi} = 5$ $TeV$ and annihilation channel is $b\bar{b}$ with annihilation rate
  $\langle \sigma v\rangle$ = $10^{-26} \mbox{cm}^3 \mbox{s}^{-1}$. The values of
  diffusion coefficient and magnetic fields are 
  $D_0 = 3 \times 10^{28} \mbox{cm}^2 \mbox{s}^{-1}$ and $B = 1$ $\mu G$, respectively.
  {\it Right panel:} Same as left panel, but for $m_{\chi} = 300$ $GeV$.}
  \label{nu_S_profile}
\end{figure*}

The effects of diffusion coefficient and magnetic
field on the final radio flux $S_{\nu}$ can be seen from
Figure \ref{nu_S_B_D}. The radio flux corresponding to higher diffusion coefficient
($D_0 = 10^{30} \mbox{cm}^2 \mbox{s}^{-1}$)
will be suppressed compared to the lower diffusion case 
($D_0 = 3 \times 10^{28} \mbox{cm}^2 \mbox{s}^{-1}$) due to the escape of a large number
of $e^{\pm}$ from the stellar object, as discussed
above. This suppression is slightly larger in the lower
frequency region, since the 
number density $\frac{dn}{dE}$ decreases more in the low
energy regime (see Figure \ref{total}). 
For, a constant $D_0$, say, $3 \times 10^{28} \mbox{cm}^2 \mbox{s}^{-1}$, a relatively lower value of $B$ (0.1 $\mu G$) reduces the radio flux by about an
order of magnitude in all frequency ranges. This is solely due to the fact that, although $\frac{dn}{dE}$ has similar values for different values of $B$ (1 and 0.1 $\mu G$), the synchrotron power spectrum $P_{Synch}$ decreases with decrease in $B$ (as evident from Figure \ref{P_E}).
We further emphasise that, though this analysis assumes an NFW profile for Draco, the choice of other profiles such as Burkert \cite{Colafrancesco:2006he,Burkert:1995yz} 
or Diemand et al. (2005) \cite{Diemand:2005wv}
(hereafter D05) \cite{Colafrancesco:2006he} keep 
our predictions similar, as can be seen in Figure \ref{nu_S_profile}.

\section{Heavy DM particles}\label{Effect of Heavy DM}

The radio flux obtained in terms of DM annihilation from a dSph crucially depends
on the source function 
$Q_e (E,r)$ (equation \ref{source_function}).
Let us try to explain why
one can get higher radio flux (obtained via $dn/dE$ through integration
of $Q_e (E,r)$) for higher DM masses in some cases \cite{Kar:2019mcq}. 
For this to happen, $Q_e$ corresponding to energetic $e^{\pm}$ is intuitively expected to go up for higher $m_\chi$. 
One may consider contributions of several components in the expression of $Q_e (E,r)$:

\begin{itemize}

\item $\langle \sigma v\rangle$: For a trans-TeV thermal DM $\chi$, the
factors affecting $\langle \sigma v\rangle$ are its mass ($m_{\chi}$) and the
effective couplings to SM particle pairs. While we are concerned here with
the annihilation cross sections for $\chi$ within a dSph, one also needs consistency with
the observed relic density \cite{Ade:2013zuv}. In general, the
expression for the relic density ($\Omega h^2$) indicates inverse proportionality
to $\langle \sigma v\rangle$, and $\frac{1}{m_{\chi}^2}$ occurs as the flux
factor in the denominator of the 
latter \cite{Gondolo:1990dk}. 
Thus a factor of $m_{\chi}^2$ occurs in the numerator.
Therefore, $m_{\chi} \gtrsim 1$ $TeV$ may make the relic density 
unacceptably large. One way to alleviate this is to ensure the possibility of
resonant annihilation \cite{Gondolo:1990dk}, as is possible in the MSSM through the participation of a pseudoscalar
of appropriate mass \cite{Gilmore:2007aq}. 
This also serves to enhance the observed radio flux, as we shall see later.
In addition, the annihilation of $\chi$ in the early universe
may involve channels other than those in a dSph, through co-annihilation with particles
spaced closely with it \cite{Gilmore:2007aq}. Once either of the above mechanisms is effective
for the relevant particle spectrum, a trans-TeV DM particle remains consistent with
all data including the relic density. This ensured, the pair-annihilation of a trans-TeV
DM particle in dSph's enable the production of relatively energetic $e^{\pm}$ pairs
which in turn reinforce the radio signals in the desired frequency range.

\item $N_{pairs} (r)
\left(= \frac{\rho^2_{\chi}(r)}{2 m_{\chi}^2}  \right)$: 
The numerator is supplied by observation.
For higher $m_\chi$, the denominator suppresses the number density of DM in
the dSph. Thus this term tends to bring down the flux for higher DM masses.

\item $\frac{dN^e_f(E)}{dE} B_f$: This has to have a compensatory effect for higher mass
if the suppression caused by the previously mentioned term has to be overcome.
Of course, the branching fractions $B_f$ in different channels have a role.
${dN^e/dE}$ is often (though 
not always) higher for higher DM masses. This feature
is universal at higher energies. 
This is basically responsible for
a profusion of higher energy electrons produced via cascades,
after annihilation in any channel has been taken place. More
discussion will follow on this point later.
It should be noted that  ${dN^e/dE}$ represents the probability that an
electron (positron) produced from the annihilation of {\em one pair} of
DM. This gets multiplied by the available number density of DM. Thus,
for comparable value of $B_f$ in two benchmark scenarios, higher ${dN^e/dE}$
along with higher $\langle \sigma v\rangle$ in the dSph (when that
is indeed the case), can cause enhancement of $Q_e (E,r)$ for higher $m_{\chi}$.

\end{itemize}

Finally, the available electron energy distribution $\frac{dn}{dE}$ (which, convoluted with
the synchrotron power spectrum, will yield the radio flux (i.e. equations \ref{j_Synch} and \ref{S_nu})), is obtained
by solving equation \ref{transport1} which has a generic solution 
(equation \ref{solution_transport1}).
Therefore, higher values
of the source function ($Q_e (E,r)$) can produce higher
radio flux at higher $m_\chi$.

As mentioned above, the annihilation of heavier DM particles will produce more 
energetic $e^{\pm}$, as can be seen by comparing the upper-left and 
upper-right panels of 
Figure \ref{Q_E} where we have
shown $dN^e/dE$ in different 
channels ($b\bar{b}$, $\tau^+\tau^-$, $W^+W^-$, $t\bar{t}$). 
The lower panel
shows the comparison of this $e^{\pm}$ spectrum 
for two DM masses (300 $GeV$ and 5 $TeV$) arising
from $b\bar{b}$ and
$\tau^+\tau^-$ annihilation channels. 
Note that, the values of $dN^e/dE$ for $m_\chi = 300$ $GeV$ and
5 $TeV$ are differently ordered for the $\tau^+\tau^-$ and $b\bar{b}$
annihilation channels. 
The spectrum for $b\bar{b}$ channel is the steepest
one, while $\tau^+\tau^-$ is the flattest. 
In fact, $dN^e/dE$ for the $b\bar{b}$
channel is governed mostly by charged pion ($\pi^{\pm}$) decay
at various stages of cascade. 
For the $\tau^+\tau^-$ channel, on the other hand,
there is a relative dominance of `prompt' electrons.
There remains a difference in the degree of degradation in the two cases. Such degradation
is reflected more in the low-energy part of the spectrum, and
leads to different energy distributions, which in turn is dependent 
on $m_{\chi}$ via the energy of the decaying $b$ or $\tau$.
The reader is referred to see references \cite{Colafrancesco:2000zv,1988ApJ...325...16R,PhysRevD.43.1774} 
for more detailed explanations.

\begin{figure*}[ht!]
\centering
  \includegraphics[height=0.36\textwidth, angle=0]{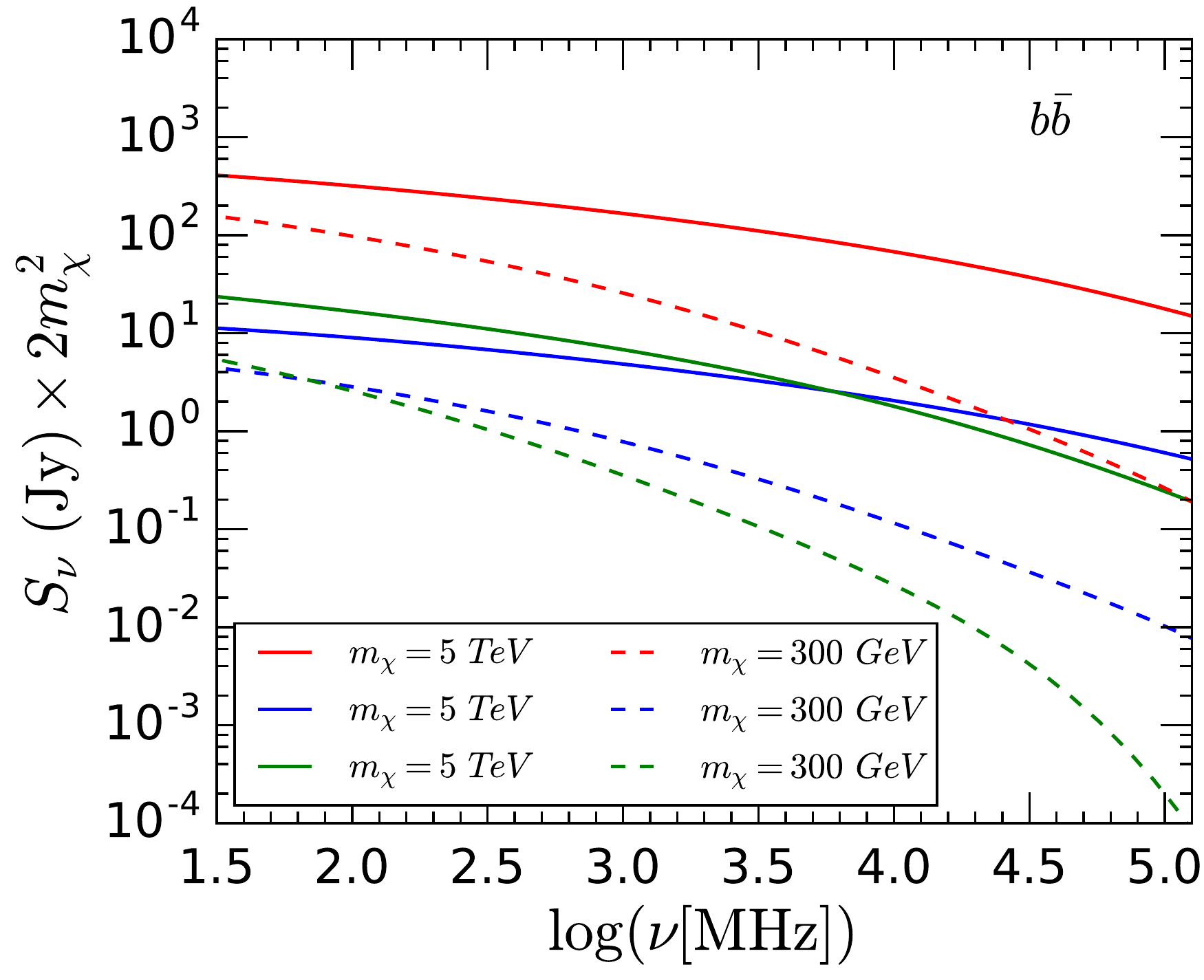}\hspace{2mm}
  \includegraphics[height=0.36\textwidth, angle=0]{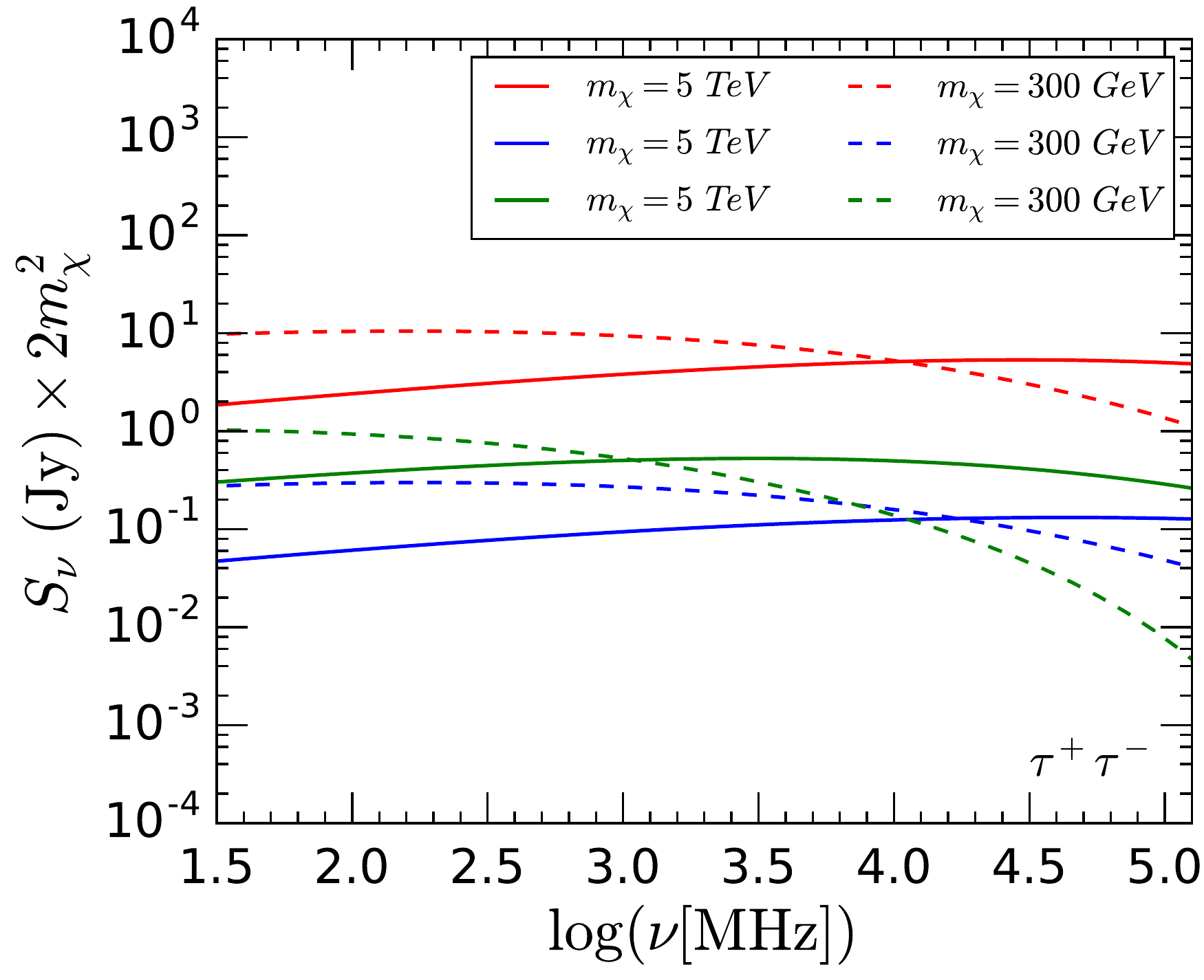}
  \caption{$S_{\nu} (Jy) \times 2 m_{\chi}^2$ vs. frequency ($\nu$) plot for two DM masses,
  300 $GeV$ (dashed lines) and 5 $TeV$ (solid lines) in two different annihilation 
  channels, $b\bar{b}$ ({\it left panel}) and $\tau^+\tau^-$ ({\it right panel}). 
  Here three different sets of astrophysical parameters have been considered, 
  $D_0 = 3 \times 10^{28} \mbox{cm}^2 \mbox{s}^{-1}$, $B = 1$ $\mu G$ (red);
  $D_0 = 10^{30} \mbox{cm}^2 \mbox{s}^{-1}$, $B = 1$ $\mu G$ (blue)
  and
  $D_0 = 3 \times 10^{28} \mbox{cm}^2 \mbox{s}^{-1}$, $B = 0.1$ $\mu G$ (green). 
  Annihilation rate
  $\langle \sigma v\rangle$ = $10^{-26} \mbox{cm}^3 \mbox{s}^{-1}$ for all cases.}
\label{nu_S_compare}  
\end{figure*}

The presence of energetic $e^{\pm}$ in 
greater abundance, which is a consequence of higher 
$m_{\chi}$,
enhances the resulting radio signal obtained through equation
\ref{solution_transport1} and \ref{j_Synch}. 
As a result of this, the final radio flux $S_{\nu}$ (equation \ref{S_nu})
gets positive contribution for higher DM masses compared to that for relatively lower
masses for most of the annihilation channels (especially
$b\bar{b}$, $W^+W^-$, and $t\bar{t}$). This can
easily be seen if one compares the quantity 
$S_{\nu} \times 2 m_{\chi}^2$ 
(i.e. removing the effect of the multiplicative factor
$\frac{1}{m_{\chi}^2}$ present in the expression of $N_{pairs}$) 
for two DM masses, 300 $GeV$ and 5 $TeV$,
in the $b\bar{b}$ annihilation channel
(left panel of Figure \ref{nu_S_compare}). If one chooses
same annihilation rate (e.g. 
$\langle \sigma v\rangle$ = $10^{-26} \mbox{cm}^3 \mbox{s}^{-1}$), 
the higher DM mass will give
higher $S_{\nu} \times 2 m_{\chi}^2$ (mainly in the high frequency region)
for this channel. We have explicitly shown this for different choices of astrophysical
parameters as indicated by different colors in the figure.

In scenarios where $\langle \sigma v\rangle$ 
corresponding to a particular $m_{\chi}$ is calculated from 
the dynamics of the model, it can happen that for
some particular benchmark the annihilation
rate ($\langle \sigma v\rangle$) is higher for larger $m_{\chi}$ (as mentioned 
earlier and will be discussed further
in the next section). In those cases
larger $\langle \sigma v\rangle$ can at least partially compensate
the effect of $\frac{1}{m_{\chi}^2}$
suppression (coming from $N_{pairs}$) for higher $m_{\chi}$. 
Thus one can get higher radio fluxes ($S_{\nu}$)
for larger DM masses compared to the smaller one.

For the $\tau^+\tau^-$ channel the situation is 
somewhat different. 
As we have already seen from the lower panel of Figure \ref{Q_E}, 
there is a large degradation of the 
$e^{\pm}$ flux ($dN^e/dE$) in this channel  
in the low energy region for higher $m_{\chi}$. 
This degradation in the source spectrum for higher $m_{\chi}$ will
continue to present in the equilibrium distribution $\frac{dn}{dE}$. 
After folding this density distribution with power spectrum (see equation \ref{j_Synch}),
the final frequency distribution will be 
suppressed for higher $m_{\chi}$, mainly in the low frequency range.
In the higher frequency range, flux is still high for
higher $m_{\chi}$ similar to that in other annihilation channel (e.g. $b\bar{b}$),
because higher $m_{\chi}$ corresponds to higher electron
distribution in the high energy range.
All these phenomena are evident from the red curves (solid and dashed)
of right panel in Figure \ref{nu_S_compare}. 
Note that, these curves are for the choice of astrophysical parameters 
$D_0 = 3 \times 10^{28} \mbox{cm}^2 \mbox{s}^{-1}$, $B = 1$ $\mu G$.
If one decreases the magnetic field $B$
to 0.1 $\mu G$ or increases the diffusion coefficient $D_0$
to $10^{30} \mbox{cm}^2 \mbox{s}^{-1}$, 
the corresponding effect can be seen from the green and
blue curves respectively.

To summarise, one can expect high radio flux for trans-TeV dark matter 
annihilation from a dSph, based on the following considerations:

\begin{itemize}

\item The high-mass DM candidate has to be consistent with relic density, 
for which $\langle \sigma v\rangle$ at freeze-out is the relevant quantity.

\item Such sizable $\langle \sigma v\rangle$ can still be inadequate in 
maintaining the observed relic density. Co-annihilation
channels may need to be available in these cases, 
though such co-annihilation does not
occur in a dSph.
This in turn may necessitate a somewhat compressed trans-TeV spectrum.

\item The high mass DM candidate should have appropriate annihilation
channels which retain a higher population of $e^{\pm}$. 
This not only off sets the suppression due to large $m_{\chi}$ but
also enhances through the energy loss term ($b(E)$) the $e^{\pm}$
density at energies low enough to contribute the radio observable
at the SKA \cite{SKA}.

\item Higher magnetic fields will be more effective in producing 
synchrotron radiation
by compensating the suppression caused by a large galactic diffusion coefficient. As we have checked
through explicit calculation, this happens even after accounting for electromagnetic 
energy loss of the $e^{\pm}$ through synchrotron effects.

\end{itemize}


\section{Detectability Curves and Final Radio Flux}\label{Detectability Curves}

The upcoming Square Kilometre Array (SKA) radio telescope will play an important role in detecting DM
induced radio signal from dwarf spheroidal galaxies. Due to its large effective 
area and better baseline coverage, the SKA has a significantly higher surface brightness sensitivity compared
to existing radio telescopes. Since the focus of this work is on the diffuse synchrotron signal from dSphs, 
we estimate the instrument sensitivity corresponding to the surface brightness and compare with the predicted signal. 
To calculate the sensitivity for a given source dSph, we need to know the baseline distribution of the telescope. The sensitivity can also be affected by the properties of the sky around the source, e.g., whether there exists any other bright source in the field of view. 

Since the detailed design of the SKA is yet to be finalised, it is difficult to estimate the noise in the direction of a given source. For our purpose, however, it is sufficient to estimate the approximate values 
of the sensitivity using the presently accepted baseline design. In order to do so, we make use of the documents provided in the SKA website \cite{SKA}.
The calculations presented in the document allow us to compute the expected sensitivity near zenith in a direction well away from the Galactic plane for all the frequencies 
relevant to the SKA, i.e., for both SKA-MID and SKA-LOW. These include contribution from the 
antenna receiver, the spill-over and the sky background (consisting of the CMB, the galactic contribution and also the atmospheric contribution). 
Note that we have assumed that all other observational systematics, such as the presence of other point sources in the field, the effect of the primary beam, 
pointing direction-dependence of the sky temperature etc are already corrected for and hence are not included in the calculation of the thermal noise.

From the above calculation, we find that typical values of the SKA surface brightness sensitivity in the
frequency range 50 MHz -- 50 GHz for 100 hours of observation time 
is $10^{-6}$ -- $10^{-7}$~Jy with a bandwidth of 300 MHz \cite{SKA,Kar:2019mcq,Colafrancesco:2015ola}. This may allow one to
observe very low intensity radio signal coming from ultra-faint dSph's.
While comparing the predicted signal with the telescope sensitivity, we have assumed that the SKA field of view is larger than the galaxy sizes 
considered here and hence all the flux from the galaxy will contribute to the detected signal.
This assumption need not be true for the SKA precursors like the Murchison Widefield Array (MWA) where
the effect of the primary beam needs to be accounted for while computing
the expected signal \cite{Kar:2019hnj}.

Till date, people do not have clear understanding about either the DM particle physics model
which governs the production of initial stage $e^{\pm}$ spectrum,
or the astrophysical parameters like galactic diffusion coefficient ($D_0$), 
magnetic field ($B$) etc. 
which are responsible for the creation of radio synchrotron flux.
Thus in this analysis, we have constrained the DM parameter space, responsible
for detecting the radio signal at SKA, for particular choices of astrophysical
parameters which are on the conservative side for
a typical dSph like Draco \cite{Colafrancesco:2006he}. On the other hand, assuming some simplified
DM model scenarios with trans-TeV DM masses,
we have estimated the limits
on the $B - D_0$ plane, required for the radio signal from Draco to be observed
at SKA.

\begin{figure*}[ht!]
\centering
\includegraphics[height=0.35\textwidth, angle=0]{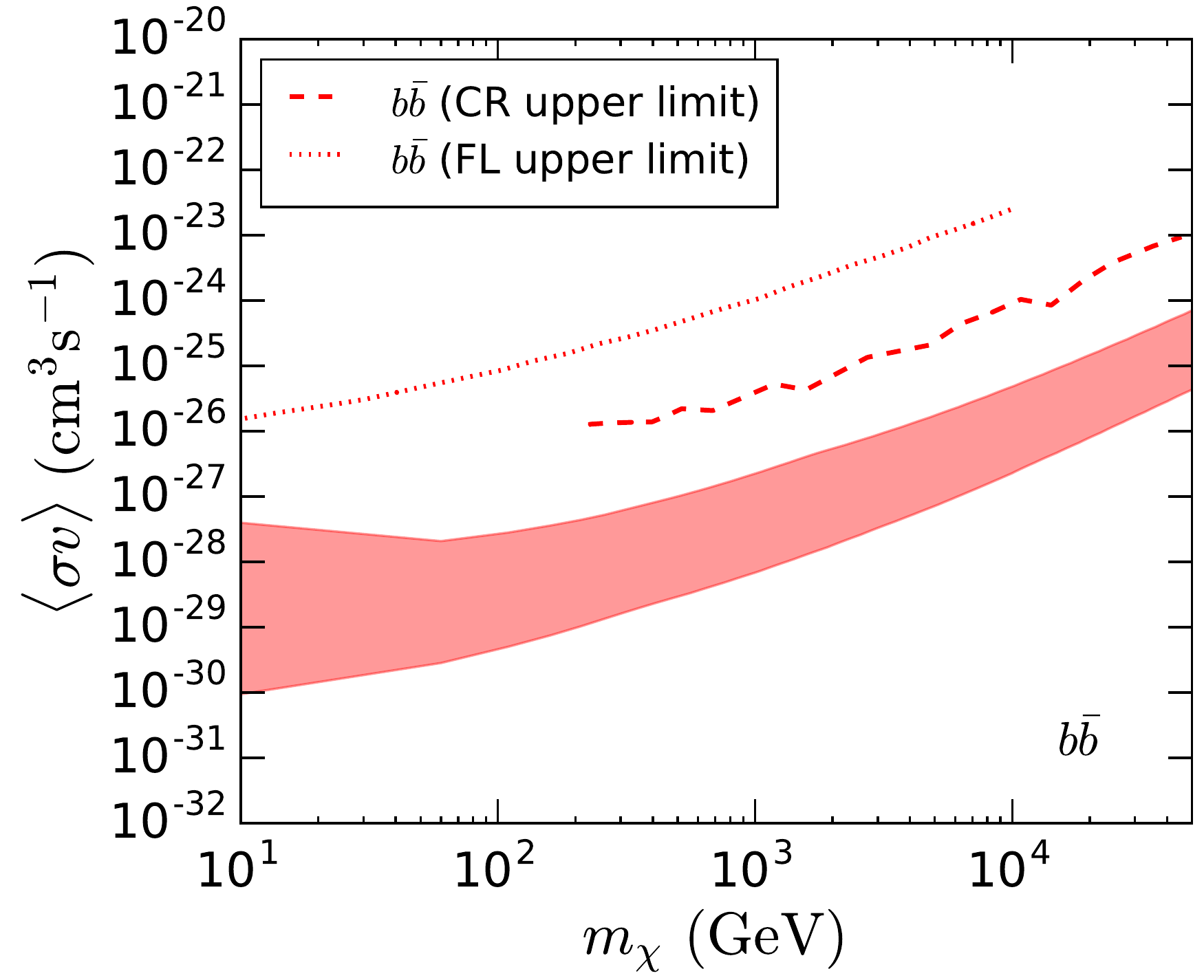}\hspace{3mm}%
\includegraphics[height=0.35\textwidth, angle=0]{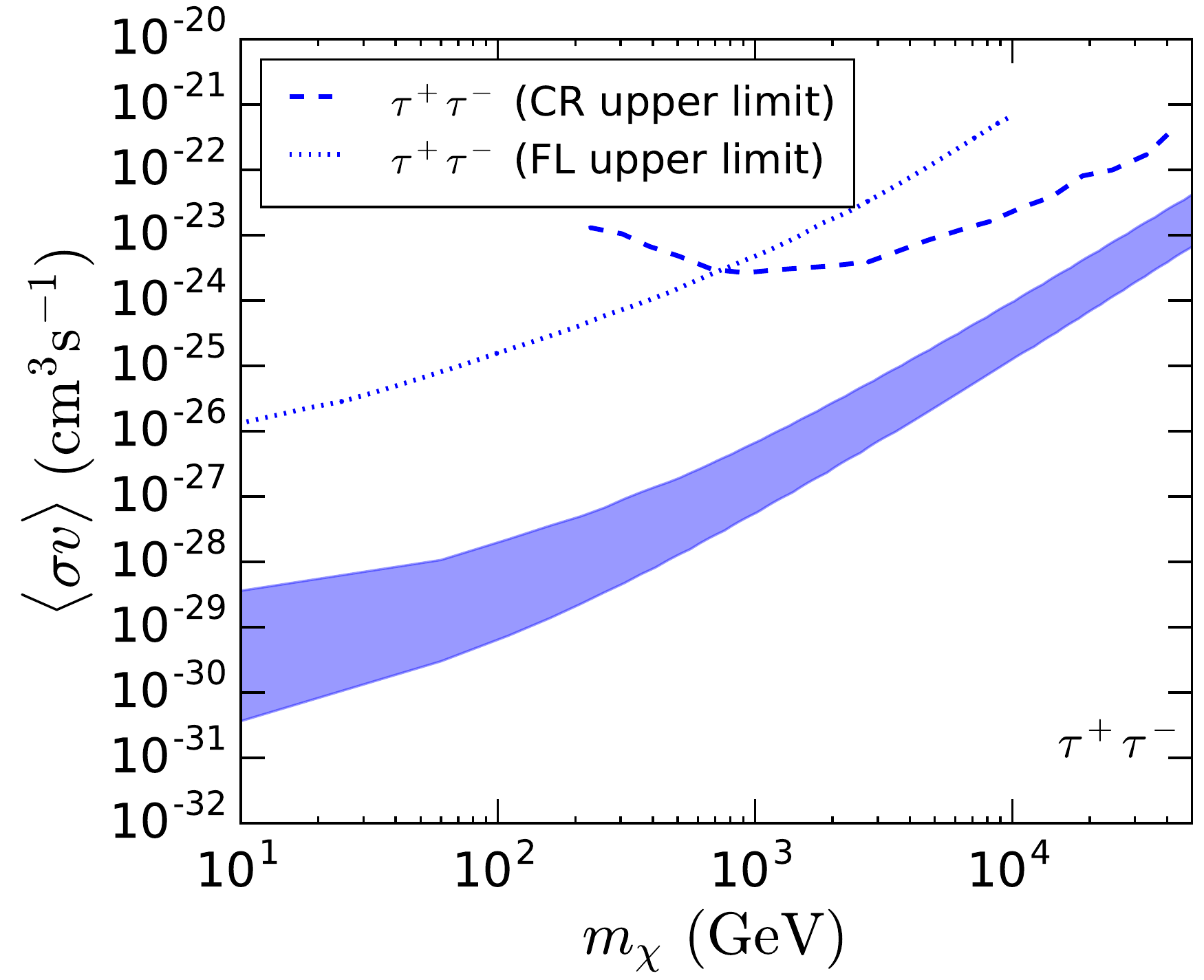}
\includegraphics[height=0.35\textwidth, angle=0]{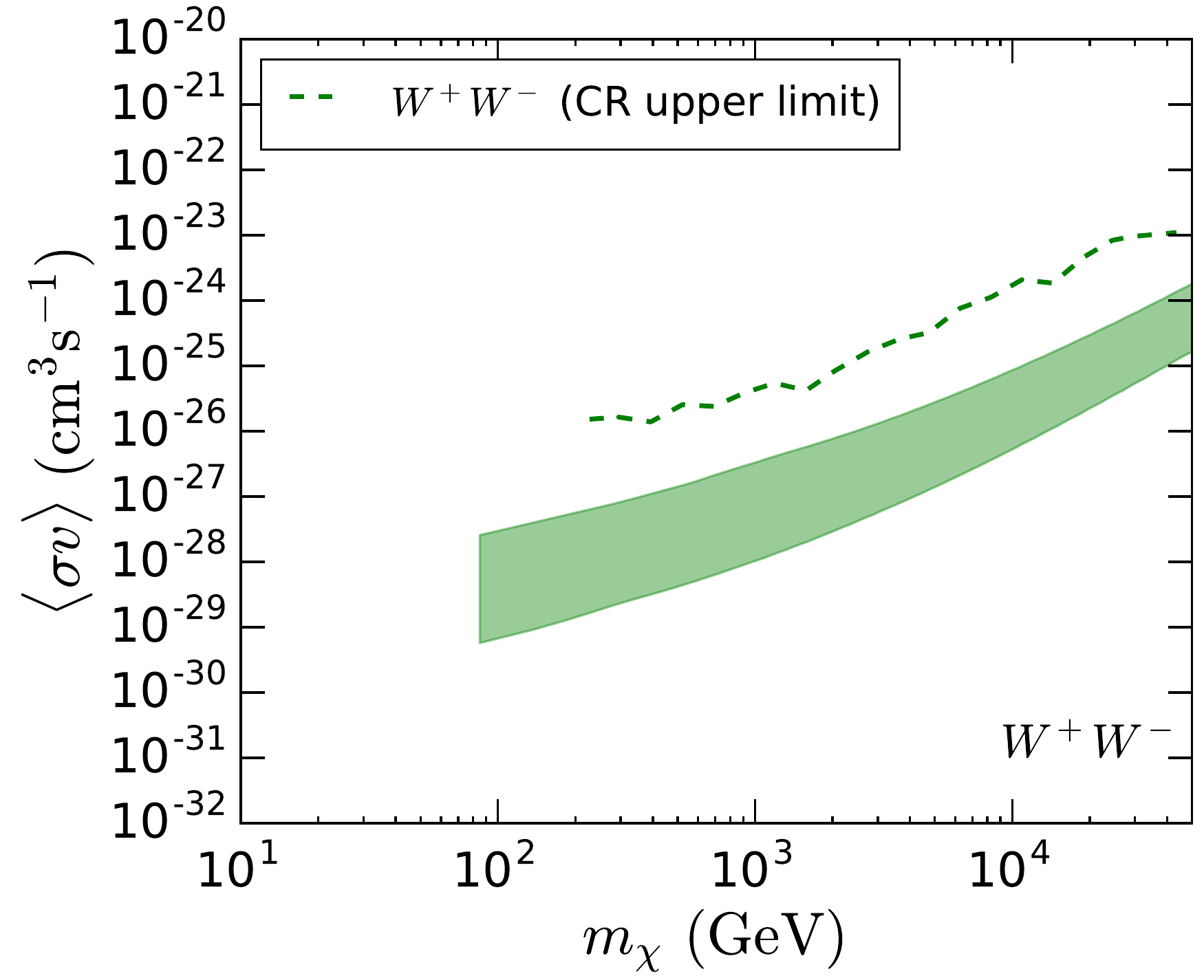}\hspace{3mm}%
\includegraphics[height=0.35\textwidth, angle=0]{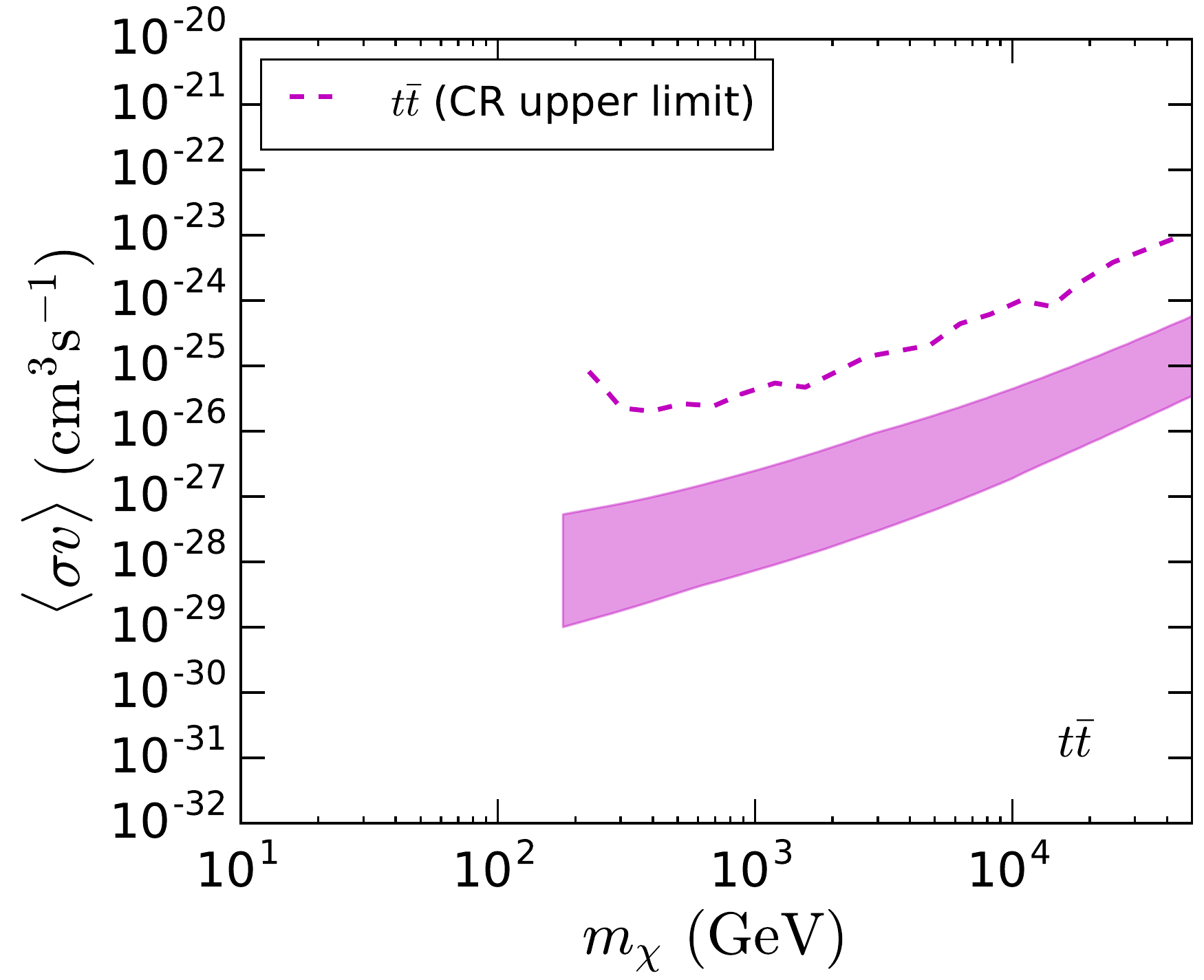}
\caption{Lower limits (colored bands) in $\langle \sigma v \rangle - m_{\chi}$ plane to observe a radio signal from 
Draco dSph at SKA with 100 hours of observation for various DM annihilation channels 
($b\bar{b}$ - {\it upper left}, $\tau^+\tau^-$ - {\it upper right}, $W^+W^-$ - {\it lower left}, $t\bar{t}$ - {\it lower right}).
For comparison, 95\% C.L. upper limits from 
Cosmic-ray (CR) antiproton observation (dashed lines) \cite{Cuoco:2017iax} 
and 6 years of Fermi-LAT (FL) data (dotted lines) \cite{PhysRevLett.115.231301} have also
been shown. 
The bands represent the variation of the magnetic field from $B = 1$ $\mu G$ (lower part of the bands) to a more conservative value
$B = 0.1$ $\mu G$ (upper part of the bands). For all cases the value of the 
diffusion coefficient ($D_0$) is $3 \times 10^{28} \mbox{cm}^2 \mbox{s}^{-1}$.}
\label{sv_mchi}
\end{figure*}

Figure \ref{sv_mchi} shows the minimum
$\langle \sigma v \rangle$ required for any $m_{\chi}$ in four different
annihilation channels ($b\bar{b}$, $\tau^+\tau^-$, $W^+W^-$, $t\bar{t}$) for 
detection of DM annihilation induced radio signal from Draco with 100 hours of observation at SKA assuming a typical band width 300 $MHz$. 
The radio signal has been taken to be detectable when the observed flux in 
any frequency bin rises three times above the noise so as to ensure that the detection is statistically significant and is not affected by spurious noise features.
We vary the DM mass over a wide range of 10 $GeV$ to 50 $TeV$, assuming 100\%
branching fraction ($B_f$) in one annihilation channel at a time.
The predictions here are for a conservative choice
of the diffusion coefficient ($D_0 = 3 \times 10^{28} \mbox{cm}^2 \mbox{s}^{-1}$) 
\cite{Colafrancesco:2006he}. 
The bands are due to the variation of the galactic magnetic field $B$
from 1 $\mu G$ (lower part of the band) to a
more conservative value 0.1 $\mu G$ (upper part of the band). As expected, the 
minimum $\langle \sigma v \rangle$ required will be larger for lower magnetic fields
as lower $B$ reduces the
the radio synchrotron frequency distribution ($S_{\nu}$). 
These limits are the detection thresholds 
for SKA to observe radio signal from Draco.
For $m_{\chi} \sim 1$ $TeV$ this limit could be as low as $\langle \sigma v \rangle \sim 3 \times 10^{-29} \mbox{cm}^3 \mbox{s}^{-1}$.
For lower values of the diffusion coefficient such as 
$D_0 = 3 \times 10^{26} \mbox{cm}^2 \mbox{s}^{-1}$ \cite{Natarajan:2015hma,Natarajan:2013dsa,Jeltema:2008ax}, the detectability
threshold band comes down even further. 
Moreover, we have not considered any halo substructure contributions
which are expected to enhance the radio flux \cite{Beck:2015rna} or lower the threshold limits even more. 
Along with these lower limits, we have also shown the model-independent upper limits (in 95\% C.L.) on $\langle \sigma v \rangle$ in various channels 
from cosmic-ray (CR) antiproton observation (dashed curves) \cite{Cuoco:2017iax} and
from 6 years of Fermi LAT (FL) $\gamma$-ray data (dotted curves) \cite{PhysRevLett.115.231301}. 
Note that the upper bounds from cosmic-ray antiproton observations are the strongest ones.
For each annihilation channel, the area
bounded by the upper and lower limits represents the region
in the $\langle \sigma v \rangle - m_{\chi}$ plane
which can be probed or constrained by SKA with 100 hours of observation. 
It is clear from the figure that even with
conservative choices of astrophysical parameters 
for $m_{\chi} \approx 50$ $TeV$, there are significantly large regions of the parameter space 
which can be probed in SKA. In such extreme cases,
it is of course necessary to have a dark sector that allows high co-annihilation
rates, so that the observed relic density bound is not excluded.

\begin{figure}[ht!]
\centering
\includegraphics[height=0.36\textwidth, angle=0]{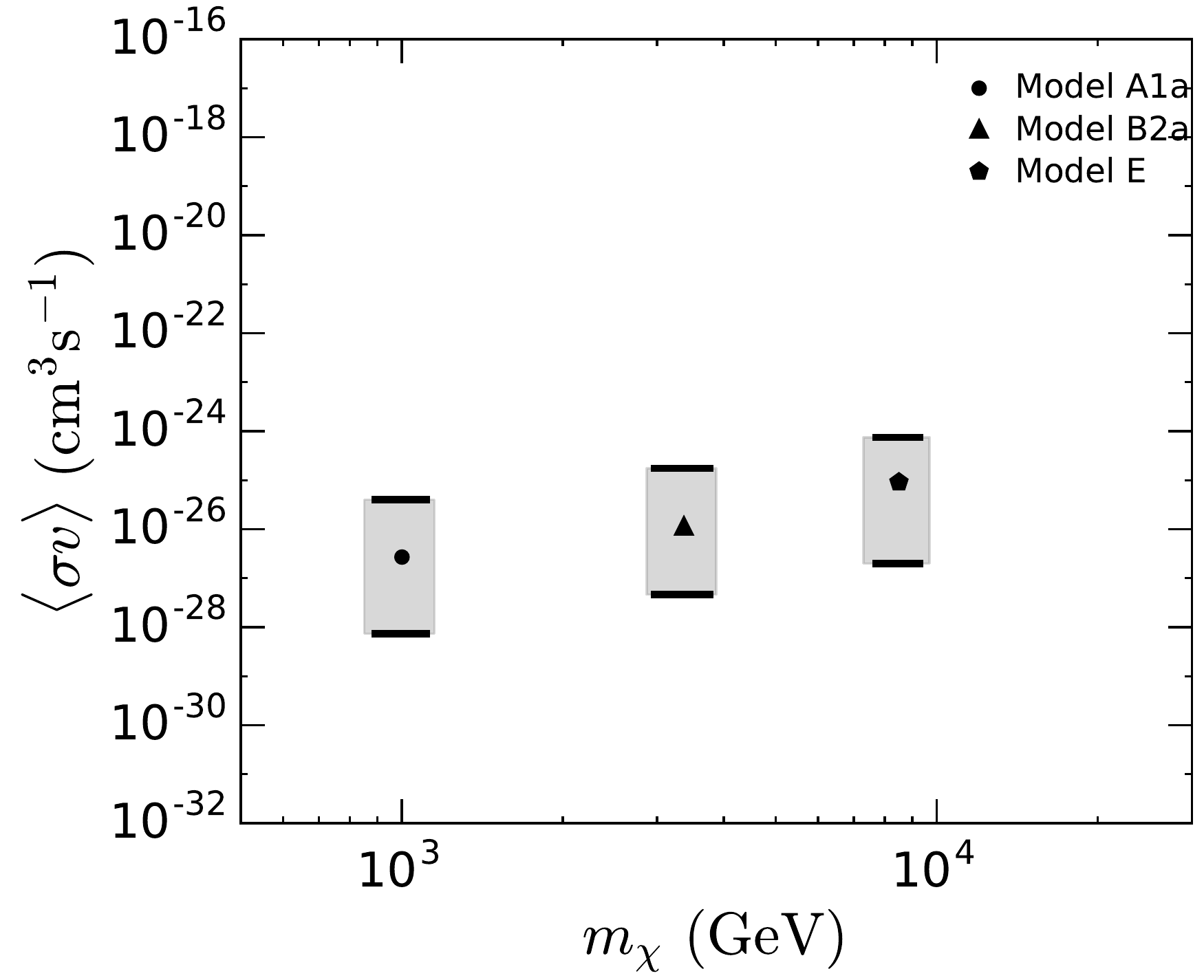}
\caption{Location of various MSSM benchmark points
(Model A1a, B2a, E; listed in table \ref{Table_benchmark}) from reference \cite{Kar:2019mcq} in the $\langle \sigma v \rangle - m_{\chi}$ plane.
The upper bars represent the 95\% C.L. upper limits on $\langle \sigma v \rangle$
corresponding to these benchmark points from 
cosmic-ray (CR) antiproton observation \cite{Cuoco:2017iax}. The lower bars show the minimum
$\langle \sigma v \rangle$ required 
for those benchmark points for the
observation of radio flux from Draco dSph at SKA with
100 hrs of observations. The diffusion coefficient
and the magnetic field have been assumed as, $D_0 = 3 \times 10^{28} \mbox{cm}^2 \mbox{s}^{-1}$
and $B = 1$ $\mu G$ respectively.}
\label{sv_mchi_benchmark}
\end{figure}

In the context of model-dependent analysis, three benchmark points,
named as Model A1a, B2a and E, from our earlier work \cite{Kar:2019mcq} have been considered here.
These benchmarks correspond to the minimal supersymmetric standard model (MSSM) 
scenario where the lightest neutralino ($\chi^0_1$) is the DM candidate ($\chi$).
Table \ref{Table_benchmark} contains the 
possible annihilation channels with branching fractions, DM masses ($m_{\chi^0_1}$) 
and annihilation rates ($\langle \sigma v \rangle$) calculated in those 
benchmark points. All these quantities have been calculated using publicly 
available package micrOMEGAs \cite{micrOMEGAs,Belanger:2001fz}.
Masses of neutralino and all other supersymmetric
particles in these three cases
are in the trans-TeV range.   
All of these benchmarks produce relic densities within the expected 
observational limits \cite{Ade:2013zuv,Harz:2016dql,Klasen:2016qyz}
and satisfy constraints coming from direct DM searches \cite{Aprile:2017iyp,PhysRevLett.118.251302},
collider study \cite{Aaboud:2017sjh}, lightest neutral 
Higgs mass measurements \cite{Allanach:2004rh} and other experiments \cite{Aaij:2017vad,Sandilya:2017mkb}. 
These benchmarks have been discussed in further detailed in reference \cite{Kar:2019mcq}. 
Figure \ref{sv_mchi_benchmark} shows 
the detectability of these benchmarks for Draco dSph
in SKA after 100 hours of observations. The predicted flux in each case falls
within the area which is still allowed by the data
and is at least 3 times above the observation threshold of SKA.
The upper limits here are taken from 
cosmic-ray antiproton observation at 95\% C.L \cite{Cuoco:2017iax}
and the lower limits have been calculated for $D_0 = 3 \times 10^{28} \mbox{cm}^2 \mbox{s}^{-1}$ and 
$B = 1$ $\mu G$. Along with these we have indicated the total
$\langle \sigma v \rangle$ (listed in third column of table \ref{Table_benchmark}) by
different points for each benchmark. It should be noted that
each such benchmark allows more than one annihilation channel. 
Consistently, the cross-sections are obtained by taking a sum over different channels appropriately weighted
by the branching ratios:

\begin{equation}
\langle \sigma v \rangle = \sum_{f} \langle \sigma v \rangle_f  B_f
\label{sv_up_low}
\end{equation}

It is clear that all these high mass cases,
which are allowed by cosmic-ray antiproton data, can easily be probed in
SKA with 100 hours of observations. This is due to the following reason:

\begin{itemize}

\item $\langle \sigma v \rangle$ is more effective in offsetting 
$\frac{1}{m_{\chi}^2}$ suppression (present in the expression for number
density of DM pair; equation \ref{source_function}) partially, though not fully.

\item A greater abundance of high-energy
$e^{\pm}$ is created by high $m_{\chi}$. This, via electromagnetic
energy loss driven by the term proportional to $b(E)$ in equation 
\ref{transport1}, generates a bigger flux of radio synchrotron
emission, as evinced in the expression for $J_{Synch} (\nu, r)$ (equation
\ref{j_Synch}).

\item The cases where one predicts more intense radio flux for
higher $m_{\chi}$ have DM annihilation mostly in the $b\bar{b}$ channels,
as against the $\tau^+\tau^-$ channel. The corresponding cascade branching 
ratios as well as the three-body decay matrix elements and their
energy integration limits are responsible for bigger radio flux.\footnote{Such effects 
can in principle be also expected if the 
$t\bar{t}, W^+W^-$ branching ratios dominates.}

\end{itemize}

\begin{table*}[ht!]
\begin{center}
\def\arraystretch{1.5}%
\begin{tabular}{|c|c|c|c|c|c|c|c|}
\hline 
Model & annihilation channel ($B_f$) & $m_{\chi^0_1}$ & $\langle \sigma v \rangle$\\
      &                              & ($GeV$)          & ($10^{-26} \mbox{cm}^3 \mbox{s}^{-1}$)\\ 
\hline
A1a & $b\bar{b}(85\%)$, $\tau^+\tau^-(14\%)$ & 1000.6 & 0.27\\
\hline
B2a & $b\bar{b}(76\%)$, $\tau^+\tau^-(15\%)$, $W^+W^-(3\%)$, $t\bar{t}(3\%)$, $ZZ(2.8\%)$ & 3368.0 & 1.19\\     
\hline
E & $b\bar{b}(79.1\%)$, $\tau^+\tau^-(18.3\%)$, $t\bar{t}(2.5\%)$ & 8498.0 & 9.12\\
\hline
\end{tabular}
\caption{Lightest neutralino mass ($m_{\chi^0_1}$) and its 
pair annihilation rate ($\langle \sigma v \rangle$) inside a dSph along with
branching fractions ($B_f$ in percentage) in different annihilation channels for the selected MSSM
benchmark points from \cite{Kar:2019mcq}. Lightest neutralino is the DM candidate 
($m_{\chi^0_1} = m_{\chi}$).}
\label{Table_benchmark}
\end{center}
\end{table*}

\begin{figure*}[ht!]
\centering
\includegraphics[height=0.35\textwidth, angle=0]{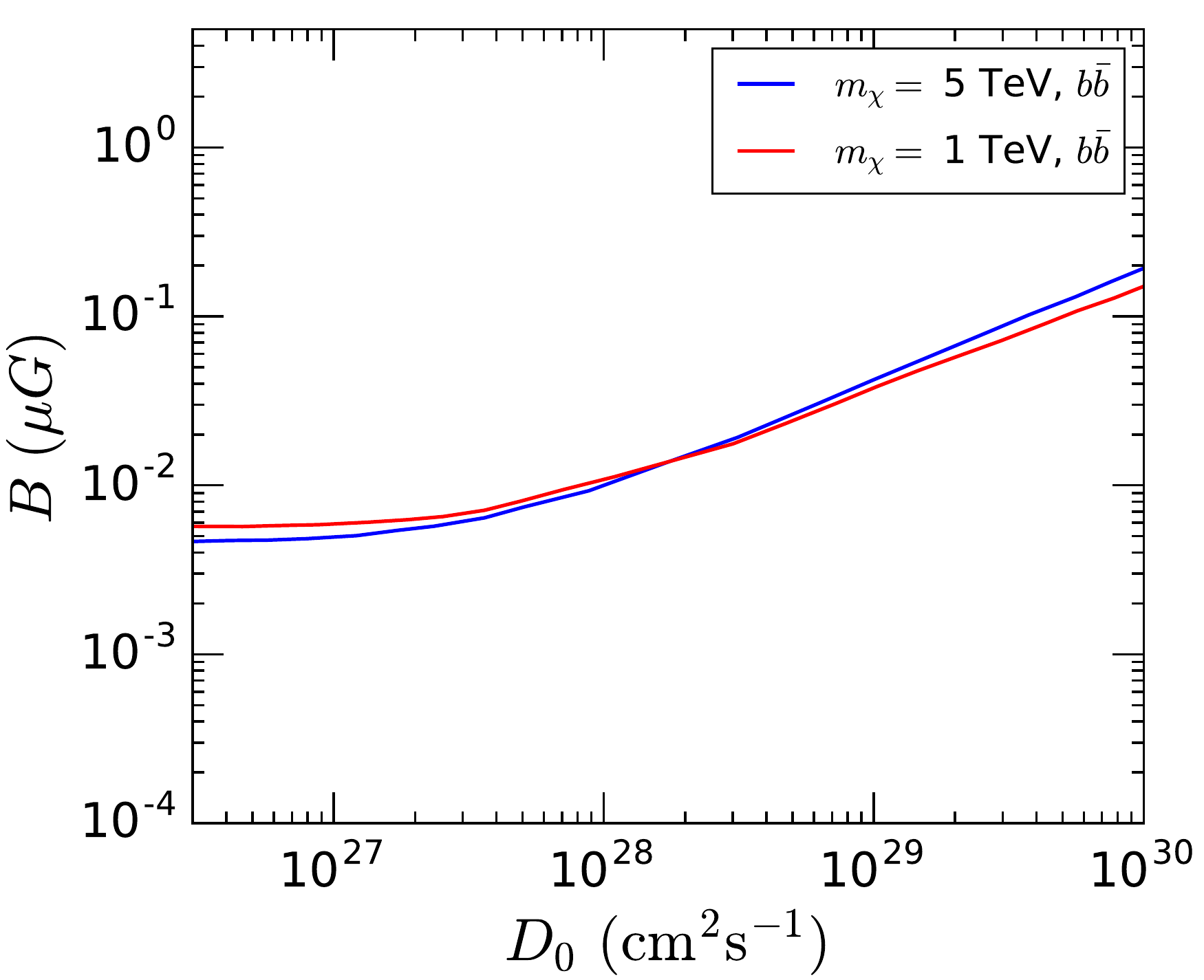}\hspace{3mm}%
\includegraphics[height=0.35\textwidth, angle=0]{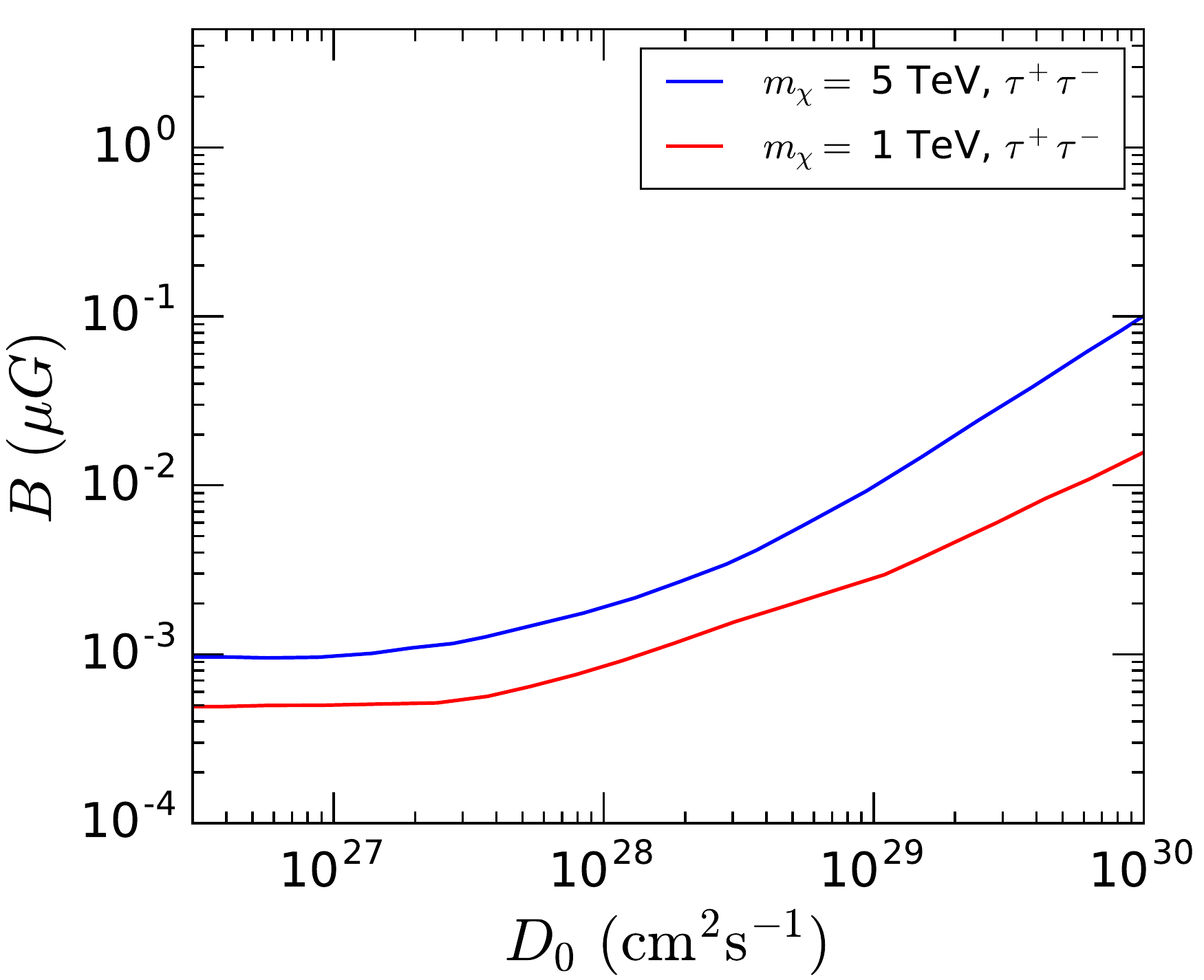}
\caption{Limits in the $B - D_0$ plane to observe a radio signal from Draco 
dSph with 100 hours of observation at SKA for $m_{\chi} = 5$ $TeV$ (blue curves) and 1 $TeV$ (red curves). Annihilation channels are $b\bar{b}$ 
(left panel) and $\tau^+\tau^-$ (right panel). The DM annihilation 
rate ($\langle \sigma v \rangle$) in each cases has been assumed to be the 95\% C.L. upper limit as obtained from cosmic-ray antiproton observation \cite{Cuoco:2017iax}.}
\label{B_D0}
\end{figure*}

Figure \ref{sv_mchi} shows how the detectability threshold at the SKA 
varies with the galactic magnetic field $B$, for a fixed value of $D_0$. 
However, the values of these two parameters for the dSphs are quite uncertain. 
In view of this, it is also important
to find out which regions in the astrophysical parameter space are within the scope of 
the SKA with 100 hours of observation, when both $B$ and $D_0$ vary over substantial ranges.
The detectability of the signal in the $B-D_0$ space is shown in Figure \ref{B_D0} for some 
illustrative DM scenarios.
These correspond to DM masses of 1 (red curve) and 5 (blue curve) $TeV$.
Cases where the dominant annihilation channel is $b\bar{b}$ are shown
in the left panel, while the curves on the right panel capture the corresponding situations with 
$\tau^+\tau^-$ as the main channel.
The solid line of each color corresponds to the maximum value of $\langle \sigma v \rangle$ for the chosen 
$m_{\chi}$, consistent with the cosmic-ray upper limit.
All points above and on the left of the curve correspond to the combinations of $B$ and $D_0$ that make the radio signals detectable over 100 hours of observations with the SKA.
Points for higher $B$ in this region correspond to models that are detectable even with lower values of 
$\langle \sigma v \rangle$. Similarly, the detectable models having progressively lower 
$\langle \sigma v \rangle$ are arrived at, as one moves to lower $D_0$ for a fixed 
value of $B$. {\it It is evident from both the $b\bar{b}$ and $\tau^+\tau^-$ channels
(chosen for illustrations in Figure \ref{B_D0}) that consistent values of 
$\langle \sigma v \rangle$ can lead to detectability with 100 hours at SKA, for $B \gtrsim 5 \times 10^{-4}$~$\mu$G for $D_0 \sim 10^{27} \mbox{cm}^2 \mbox{s}^{-1}$. On the other hand, for larger values of $D_0 \sim 10^{30} \mbox{cm}^2 \mbox{s}^{-1}$, we need $B \gtrsim 10^{-2}~\mu$G for the signal to be detectable.}

\begin{figure}[ht!]
\centering
\includegraphics[height=0.36\textwidth, angle=0]{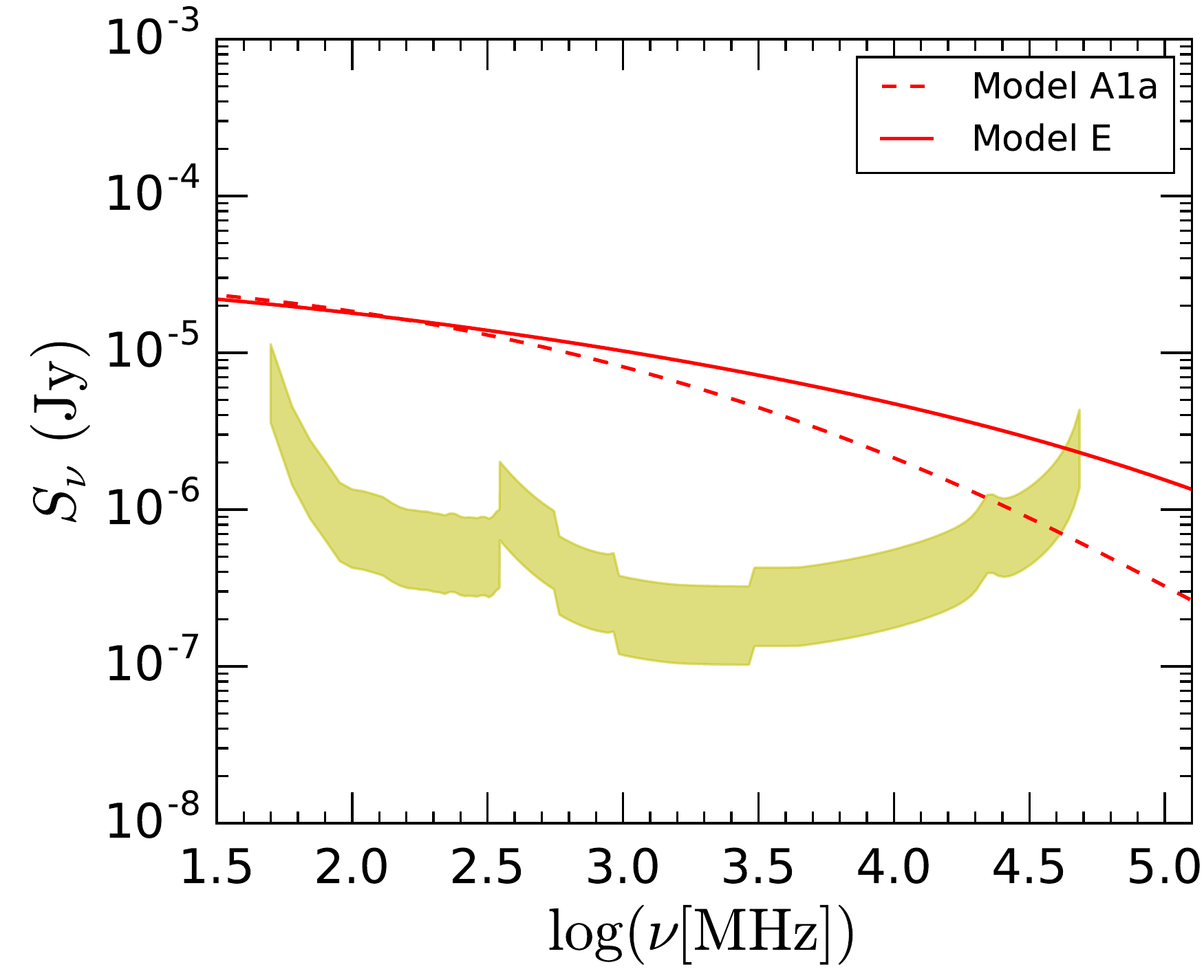}
\caption{Radio synchrotron flux ($S_{\nu}$) vs. frequency ($\nu$) 
plot for Draco dSph using two MSSM benchmark
points A1a and E from table \ref{Table_benchmark}. The yellow shaded band denotes 
the SKA sensitivity corresponding to the
variation of observation time from 10 hours (upper part of the
band) to 100 hours (lower part of the band) \cite{SKA, Kar:2019mcq}. 
Diffusion coefficient 
$D_0 = 3 \times 10^{28} \mbox{cm}^2 \mbox{s}^{-1}$ 
and magnetic field $B = 1$ $\mu G$.}
\label{nu_S_MSSM}
\end{figure}

Finally, we show in Figure \ref{nu_S_MSSM} the final 
radio fluxes in two of the MSSM benchmarks listed in table 
\ref{Table_benchmark}. For illustration
we have taken the benchmark points A1a ($m_{\chi} \sim 1$ $TeV$)
and E ($m_{\chi} \sim 8.5$ $TeV$).
The annihilation is $b\bar{b}$-dominated for both cases. 
The yellow band here represents the SKA sensitivity \cite{SKA}, with a bandwidth of 300 $MHz$.
The band is due to the variation of the observation time from 10 hours
(upper part of the band) to 100 hours (lower part of the band) \cite{SKA, Kar:2019mcq}.
The choice of the diffusion coefficient
and the magnetic field is on the conservative side 
($D_0 = 3 \times 10^{28} \mbox{cm}^2 \mbox{s}^{-1}$ 
and $B = 1$ $\mu G$). We have already shown the 
detectability of these benchmarks in Figure 
\ref{sv_mchi_benchmark}. From this figure we can see that, 
even for 10 hours of observation, these high-$m_{\chi}$ benchmarks can easily be observed in SKA over most of its frequency range.

One important feature of Figure \ref{nu_S_MSSM} is that, 
in a wide range of frequency (300 $MHz$ -- 50 $GHz$),
suitable for SKA, $S_{\nu}$ for 
very high mass case (Model E) is higher than 
that for low mass case (Model A1a). Though the DM mass in Model E
is larger, annihilation rate in this case is higher 
($\sim 9 \times 10^{-26} \mbox{cm}^3 \mbox{s}^{-1}$) 
than in Model A1a 
($\sim 3 \times 10^{-27} \mbox{cm}^3 \mbox{s}^{-1}$).
In general, $\langle \sigma v \rangle$ should have a $\frac{1}{m_{\chi}^2}$
suppression, because of the energies of the
colliding DM particles \cite{Gondolo:1990dk}. 
In spite of that, Model E (higher $m_{\chi}$) has
greater $\langle \sigma v \rangle$\footnote{higher value of $\langle \sigma v \rangle$
for higher $m_{\chi}$ is also needed to keep the
relic density under the observed limit, 
even when there is scope
of co-annihilation in the early universe.}
than Model A1a (lower $m_{\chi}$)
mainly due to the
closer proximity to a s-channel resonance mediated by CP-odd
pseudoscalar
in the annihilation process like 
$\chi^0_1 \chi^0_1 \rightarrow b \bar{b}$. 
For detailed information reader is referred to
see reference \cite{Kar:2019mcq}

As discussed earlier, The higher value 
of $\langle \sigma v \rangle$ partially offsets 
the effect of larger DM mass and consequently we get $\frac{\langle \sigma v \rangle}{m_{\chi}^2}$ (which appears in the source function; equation \ref{source_function})
for Model A1a and Model E as 2.7 and 1.26 
(in units of $10^{-33} \mbox{GeV}^{-2} \mbox{cm}^3 \mbox{s}^{-1}$), respectively,
though the latter model has larger $m_{\chi}$ ($\sim 8.5$ $TeV$) compared to 
the former ($\sim 1$ $TeV$).
Thus a 72-fold suppression due to $m_{\chi}^2$ results in a suppression just by a factor of
$\approx 2$ at the level of $\frac{\langle \sigma v \rangle}{m_{\chi}^2}$.
Also, note that the $b\bar{b}$ annihilation channel dominates for both models.
These observations, together with 
the discussion in section \ref{Effect of Heavy DM} 
and the contents of Figure
\ref{nu_S_compare}, explains a higher mass DM particle
generating higher radio flux for scenarios where the $b\bar{b}$ annihilation channel 
dominates over $\tau^+\tau^-$.

\section{Conclusion}\label{Conclusion}

The aim of this paper has been to show that radio signal arising from a high mass (trans-TeV)
WIMP DM can be detectable as radio synchrotron flux from a dSph, to be recorded by the 
upcoming SKA Telescope. We have analysed not only the particle physics aspects of DM
annihilation and subsequent cascades leading to $e^{\pm}$ pairs, but have also
included the relevant astrophysical processes the electrons/positrons pass through before 
emitting radio waves, upon acceleration by the galactic magnetic field.
We have set out to identify the mechanism whereby a trans-TeV DM candidate 
can thus be visible in radio search. 
We found that in the SKA frequency range,
enhancement of the radio flux for this case is possible mainly due to the
following reasons:

\begin{itemize}

\item Larger cross section or annihilation rate (required to maintain
relic density under the observed limit for a trans-TeV DM) 
facilitated by the dynamics of the particle physics model. This helps in compensating
the $\frac{1}{m_{\chi}^2}$ suppression due to large DM mass.

\item Presence of energetic $e^{\pm}$ in the DM annihilation spectrum
in greater abundance.
This partially reduces the $\frac{1}{m_{\chi}^2}$ suppression effect and 
on the other hand enhances, through the energy loss term, the electron/positron density
at low energies which helps to produce large radio flux. 

\item Dominance of the annihilation channel $b\bar{b}$ which yields a comparatively 
larger abundance of $e^{\pm}$ in all of the energy range
of the spectrum produced by DM annihilation.

\end{itemize}

Simultaneously, effect of various astrophysical parameters 
(e.g. $D_0$, $b(E)$, $B$) 
on the radio synchrotron flux produced from the annihilation of a trans-TeV
DM particle has been studied in detail.

Using SKA sensitivity we have drawn the limits in the 
$\langle \sigma v \rangle - m_{\chi}$ plane to observe radio flux from
Draco with 100 hours of observation.
We found that, these limits are much more stronger than the previously obtained 
bounds on $\langle \sigma v \rangle$
from Fermi-LAT $\gamma$-ray \cite{PhysRevLett.115.231301} or AMS-02 cosmic-ray antiproton observation \cite{Cuoco:2017iax}. 
Even for a conservative choice of astrophysical parameters
($D_0 = 3 \times 10^{28} \mbox{cm}^2 \mbox{s}^{-1}$ 
and $B = 1$ $\mu G$), we found that these limits can go as low as 
$\langle \sigma v \rangle \sim 3 \times 10^{-29} \mbox{cm}^3 \mbox{s}^{-1}$ for a 
$m_{\chi} \sim 1$ $TeV$. This indicates towards a large region of
WIMP parameter space which can be probed through the upcoming SKA.
Along with these we have also shown the limits in the 
$B - D_0$ plane.
We found that, for a DM mass
in the trans-TeV range, magnetic field as low as $B \sim 10^{-3}$ $\mu G$ and
diffusion coefficient as high as 
$D_0 \sim 10^{30} \mbox{cm}^2 \mbox{s}^{-1}$ are well enough to produce
radio flux above SKA sensitivity in 100 hours of observation time.

Taking the minimal supersymmetric standard model (MSSM) as an illustration, we have
shown that benchmark points with lightest
neutralino masses ($m_{\chi^0_1}$) $\sim$ 1 - 8 $TeV$, which satisfy all the constraints
from observed relic density, direct DM searches and collider searches, can
lead to detectable signals at the SKA with 100 hours observation, even with conservative choice of $B$ and $D_0$.
We have illustrated how the effects mentioned earlier can lead to a larger radio flux
for a high mass DM benchmark point compared to a low mass case, thus
establishing the credibility of the search for heavy DM
through radio observation in SKA\footnote{After submitting this paper,
we came to know of reference \cite{Cembranos:2019noa}, where a similar
study has been carried out for some different scenarios. While the broad conclusions
have some similarity, our conclusions are a little more optimistic, since we  have
predicted detectability with 100 hours of observation, with realistic benchmarks.
One of the reasons that have led to the more positive predictions is the inclusion
of the $b\bar{b}$ annihilation channel in our study.}.

\vspace{1cm}
{\bf Acknowledgements}\\

The work of AK and BM was supported by the
funding available from the Department of Atomic Energy, Government of India, for the Regional 
Centre for Accelerator-based Particle Physics (RECAPP), Harish-Chandra Research
Institute. TRC and SM acknowledge the hospitality of RECAPP during the formulation of the project.


\bibliography{biblio} 

\end{document}